\documentclass[12pt]{article}
\usepackage{graphicx} % Required for inserting images
\usepackage[left=2.5cm,top=2.5cm,right=2.5cm,bottom=2.5cm]{geometry}
\usepackage{amsmath} 
\usepackage{amssymb}
\usepackage{hyperref}
\usepackage{authblk}
\usepackage{xcolor}
\usepackage{caption}
\usepackage{ulem}

\makeatletter
\newcommand\ackname{Acknowledgements}
\if@titlepage
  \newenvironment{acknowledgements}{%
      \titlepage
      \null\vfil
      \@beginparpenalty\@lowpenalty
      \begin{center}%
        \bfseries \ackname
        \@endparpenalty\@M
      \end{center}}%
     {\par\vfil\null\endtitlepage}
\else
  
\fi
\makeatother

\title{Eclipses by Artificial Satellites \\to Measure the Angular Sizes of Stars}

\author[1,*]{Leo W.H. Fung}
\author[2]{Albert Wai Kit Lau}
\author[1]{\\Richard Massey}
\author[3,4,5,6,7,8]{George F. Smoot$^\dagger$}

\affil[1]{Institute for Computational Cosmology \& Centre for Extragalactic Astronomy, Durham University, Stockton Rd, Durham DH1 3LE, UK}
\affil[2]{Dunlap Institute for Astronomy and Astrophysics, University of Toronto, Toronto, Canada}

\affil[3]{Department of Physics, University of California, Berkeley, California, USA, \it emeritus}
\affil[4]{Institute for Advanced Study, The Hong Kong University of Science and Technology, Clear Water Bay, Kowloon, Hong Kong, \it emeritus}
\affil[5]{Energetic Cosmos Laboratory, Nazarbayev University, Astana, Kazakhstan, \it emeritus}
\affil[6]{Laboratoire APC-PCCP, Université Sorbonne Paris Cité, Université Paris Diderot, \it emeritus}
\affil[7]{Laboratoire Astroparticule et Cosmologie, Universit{\'e} de Paris, F-75013, Paris, France, \it emeritus}
\affil[8]{Donostia International Physics Center,  University of the Basque  Country  UPV/EHU,  E-48080  San  Sebastian, Spain}
\affil[*]{Corresponding author: wing.h.fung@durham.ac.uk $|$ leowhfung@gmail.com}

\newcommand\blfootnote[1]{%
  \begingroup
  \renewcommand\thefootnote{}\footnote{#1}%
  \addtocounter{footnote}{-1}%
  \endgroup
}

\begin{document}
\maketitle
\begin{abstract}
Direct measurements of the angular sizes of stars (other than our Sun) are inherently limited by telescopes' optical diffraction limit. 
The spatial diffraction limit can be overcome by using time-domain information when stars are `eclipsed' by anything moving in the foreground: artificial satellites or the Moon (known as `lunar occultation imaging').
%It is possible to circumvent the diffraction limit by transcribing time domain information to recover the sub-diffractive spatial information, as demonstrated by lunar occultation imaging technique deployed in the 2000s.
% Techniques for circumventing the diffraction limit are often based on encapsulating the spatial information in the time domain, with occultation imaging being one of the promising approaches that had been demonstrated over the last decade.
%High-speed photon counting devices have developed rapidly over the last decade. 
Here we analyse the angular resolution achievable by high-speed photon counting devices, when stars are eclipsed for several microseconds by one of the thousands of satellites now in low or mid-Earth orbit.
%in the context of microsecond occultation, during which the distant star light is temporally blocked by a fast flyby object in the foreground.
%We simulate different types of occultation, ranging from the classical lunar occultation, to the unexplored domain of microseconds ($\mu$s) occultation by artificial satellites.
We also build a full Bayesian statistical treatment for analysing simulated observations.

We show that eclipses by the moon and satellites deliver $<2$ milliarcsecond resolution for $\sim$$10^4$ stars every year, with the diameters of $\sim$$10^1 - 10^2$ stars constrained to better than $0.5$ milliarcseconds. 
The application of satellites requires an ultra-fast, photon-resolving detector, capable of continuous readout with inter-frame delay controlled within $\lesssim 1$\,$\mu$s. 
Such a fast rate can easily be handled by the SPINA instrument, which has demonstrated $8$\,ns inter-frame delay, and potentially by the Cherenkov Telescope Array.
% The angular resolution can be improved with the occultations by asteroids, at the expense of lower detection rate; in the meantime, the occultations by artificial satellites can deliver angular resolution comparable to that of the moon, making it suitable for supplementing other classical techniques.
% Several techniques for circumventing the diffraction limit have been proposed and implemented in the past decade

\end{abstract}
\blfootnote{$^\dagger$ Our good friend and mentor G. Smoot passed away prior to the submission of this manuscript, on 18 Sept 2025. }

\section{Introduction}

If only our telescopes could resolve angular scales below a milliarcsecond, we could directly measure the size of stars along their evolutionary tracks \cite{stellar-evolution-chandrasekhar-review1984,massive-star-evolution,stellar-structure-evolution-review2015}, optical emission from active galactic nuclei \cite{agn-size-problem-2025,agn-size-reverberation-map,agn-size-reveberation-map-measure}, and gravitational lensing by intermediate-mass black holes \cite{imbh-with-lensed-gw,imbh-detection-globular-lensing}.
At visible wavelengths, optical diffraction makes these ambitions infeasible for now.
Rayleigh's criterion for a static scene requires even a diffraction-limited telescope to have an aperture of diameter 100~metres.
This would be expensive, as the construction cost of a telescope scales as its aperture size cubed \cite{meinel-telescope-cost-scaling}.
If the giant telescope is on Earth, adaptive optics \cite{adaptive-optics-for-astro-review} would also be required to correct turbulence in the Earth's atmosphere each $\sim 1$~millisecond, to prevent smearing of $\sim100$~milliarcseconds \cite{adaptive-optics-review}. While 10~metre-class telescopes have thus approached the diffraction limit \cite{keck-adaptive-optics,subaru-adaptive-optics,vera-rubin-adaptive-optics}, the number of required real-time wavefront calculations scales as aperture size to the power six \cite{adaptive-optics-ELT-algo}. 
% \rjm{Check scaling with Alistair}.

`Super-resolution' techniques to circumvent Rayleigh's static limit have been developed in biological science, by controlling source illumination \cite{structured-illumination-origin-idea-1993,structured-illumination-first-implement-2000} or using time domain information \cite{STED-original-hell94,STED-review-2017,sr-sim-review,sr-sim-review-2021}.
In astrophysics, light emission from distant sources cannot be controlled, but time domain information is available. Intensity interferometry is the most ambitious technique, exploiting the bosonic statistics of photon bunching on $\sim 10^{-15}$~second scales \cite{hbt-effect-original1956}. Observed as early as 1956 \cite{stellar-interferometer-original1956}, recent measurements \cite{stellar-intensity-interferometry-2012,stellar-intensity-interferometry-magic,iqueye-intensity-interferometry} achieved angular precision of $0.542 \pm 0.018$~milliarcseconds for a star of AB magnitude $1.99$ (in $4.25$ hours with the VERITAS array of four $12$-metre telescopes designed to detect Cherenkov emission by cosmic gamma rays \cite{veritas-intensity-inteferometry}). Unfortunately, the technology to capture images on femtosecond timescales is challenging, and measurement precision drops rapidly for fainter sources, thus hindering its application to large surveys for population-level analysis.

Here we exploit time-domain information from `occultation imaging':  when a foreground object moves through our line of sight (LOS) to a background source, temporarily masking it. The foreground object must be nearby, because the effect relies on the diffraction of light around it. 
Occultation imaging using the Moon was first achieved in 1946 using a photo-multiplier tube with a cathode ray oscilloscope readout \cite{whilfford1947-first-lunar-occultation}.
Other early attempts (for example, \cite{evans-1955-lunar-occultation-analog}) utilising similar analogue setup had demonstrated milliarcsecond resolvability, and these efforts were collectively reviewed by \cite{lunar-occultation-early-review-1983}.
The method was improved with contemporary digital technologies in 1980s \cite{lunar-occultation-obs-1987,lunar-occultation-obs-1992}, and modern 8~metre class telescopes \cite{lunar-occultation-thai-telescope,eso-lunar-occultation-firstRun2008} have acquired $2$~milliarcsecond resolution for a magnitude $9.18$ star \cite{eso-lunar-occultation-2010,eso-lunar-occultation-catalog2011}.
Rare occultations by asteroids \cite{asteroid-occultation-2024-sCMOS,asteroid-occultation-obs-2009-ccd} have yielded measurement precision better than $0.1$~milliarcseconds for a magnitude $10.2$ star with the VERITAS array \cite{veritas-asteroid-occultation}.
Crucially, the temporal resolution of required observations is much slower than that of the intensity interferometry. The Moon occults a star in $\Delta t \sim 10^{-3}$~seconds, and an asteroid occults it in $\Delta t \sim 10^{-2}$--$10^{-1}$~seconds. This is accessible using  standard Complementary Metal-Oxide-Semiconductor (CMOS) \cite{asteroid-occultation-2024-sCMOS} or fast charge-coupled device (CCD) \cite{asteroid-occultation-obs-2009-ccd} detectors.

Occultation imaging relies upon the fortunate alignment of foreground objects with background stars that are sufficiently bright for measurements at high frame rates.
As demonstrated later in this work, the limiting V-band magnitude is approximately $10$, with the exact value scaling with telescope aperture size. 
The sparsity of such bright stars across the sky limits the rate of occultations by Solar system objects.
However, Low-Earth Orbit (LEO) has recently become more populated by artificial satellites, especially the Starlink constellation. 
While Starlink satellites may degrade wide-field sky surveys by leaving trails on long-exposure images \cite{starlink-astro-overview}, we shall demonstrate that their unprecedented occultation rate of stars could be useful for occultation imaging.
%In this paper, we demonstrate the effectiveness of using LEO satellites for occultation imaging.
Bound in its Keplerian orbit, the slowest \footnote{The slowest timescale is defined by the width of the first-order Fresnel fringe. 
The higher order fringe spacing (and thus fringe width) decays linearly. Consequently, resolving these fringes requires increasingly fast temporal resolution. } light-curve variation timescale due to a LEO satellite is $\sim10^{-4}$ seconds.
Sub-diffraction-limited measurements of the source can therefore be obtained by a photon counter with temporal resolution better than $\sim10^{-6}$ seconds so as to temporally resolve the fine structures of such variations. 
Silicon Photomultipliers (SiPMs) meeting this requirement are routinely deployed in Cherenkov telescopes \cite{cta-science-paper} and, in the Single-Photon Imager for Nanosecond Astrophysics (SPINA) experiment \cite{lau2020sky,lau2022development}, achieved temporal resolution $8 \times 10^{-9}$ seconds \cite{lau2023initial} in spatially-resolved imaging with less than one dark count during the period of a typical occultation event \cite{spina-noise-model}.
The conclusions of this paper are not very sensitive to instrument-specific details but, since SPINA is a viable real-world instrument, we shall use its parameters where necessary to simulate observations.

This paper is organised as follows. In \S\ref{sect: formalism}, we derive the theory of occultation imaging, define an effective angular resolution of observations, and calculate the rate of occultations by different classes of satellite. In \S\ref{sect: stats}, we present mock observations of occultation events, and validate a data analysis pipeline to infer physical measurements. In \S\ref{sect: results}, we present results on the resolution and rate of observations achievable with different classes of satellite. We conclude in \S\ref{sect: conclusion}.

\section{Formalism} \label{sect: formalism}

\begin{figure}[t]
    \centering
    \includegraphics[width=0.9\linewidth]{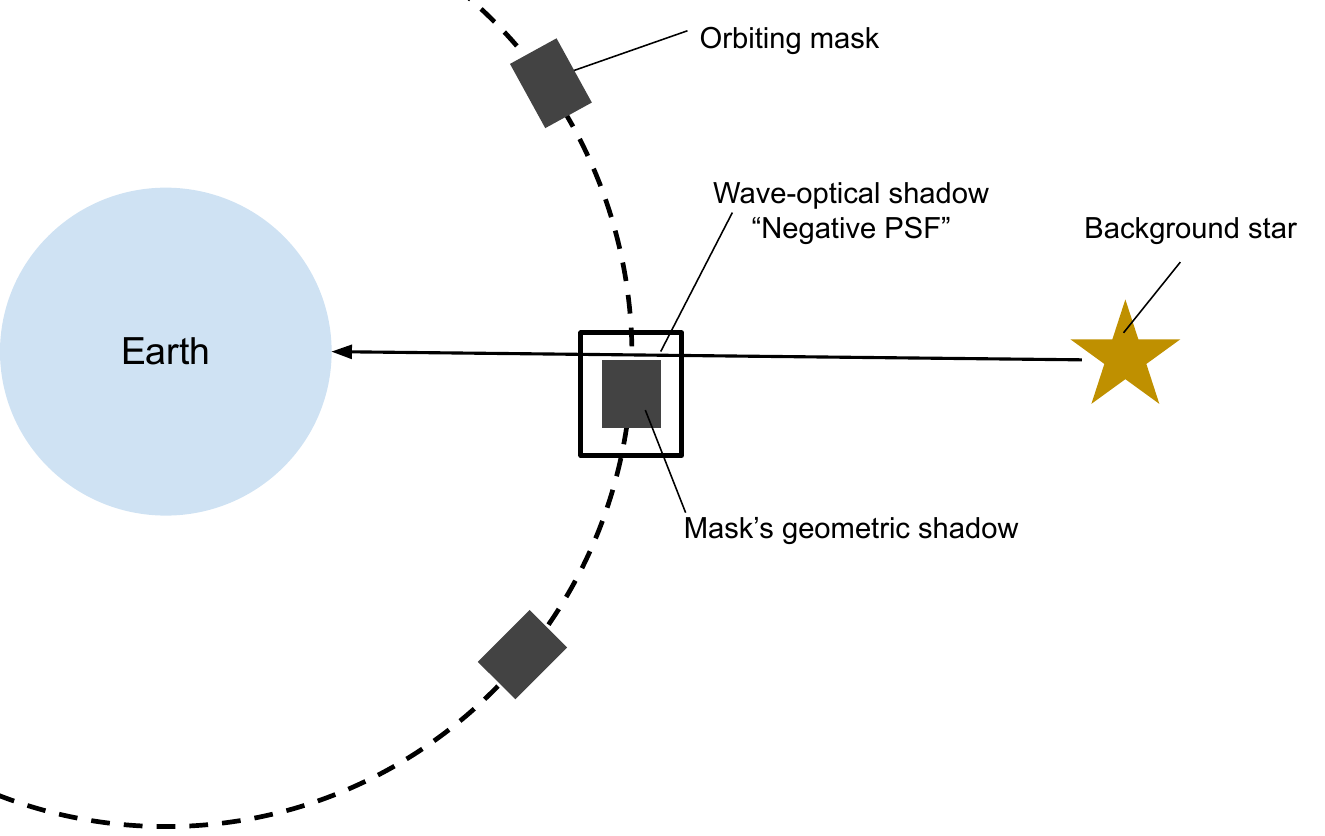}
    \caption{The geometry of occultation. During a brief flyby of an orbiting mask into the line-of-sight to a distance star, the distant starlight would be diffracted by the foreground mask before reaching telescopes on Earth. 
    This leaves a characteristic feature in the light curve.
    }
    \label{fig: occultation-cartoon}
\end{figure}

\subsection{Theory of occultation imaging}

Consider an observing setup in which light from a background point source (S) propagates towards an observer (O).
At a specific location along the line-of-sight (LOS), an intervening object (M) masks the point source. 
We denote the observer--source distance as $D_{\rm OS}$, the observer--mask distance as $D_{\rm OM}$, and the mask--source distance as $D_{\rm MS}$.
While this setup is simple, the observed effects differ drastically in the various regimes of the dimensionless ratio $D_{\rm OM}/D_{\rm MS}$.
\begin{enumerate}
    \item When $D_{\rm OM}/D_{\rm MS} \gg 1$, the geometrical optics approximation is valid. 
In this case, if the mask aligns perfectly with the LOS, all flux from the background point source is blocked. 
If the background source is larger than the mask, the total flux received is reduced by the ratio of the mask area to that of the background. 
This occurs during exoplanet transits \cite{exoplanet-transit-review}, where the timescale over which the observed flux drops to a minimum can be directly translated into the size of the mask (i.e. the exoplanet). This process of converting spatial information into observable time-domain variations allows one to bypass the telescope's angular resolution limit.
\item In this work, we focus on the opposite limit, where $D_{\rm OM}/D_{\rm MS} \ll 1$, meaning the observer is in the near-field of the mask and wave-optics effects become relevant. 
While the intuition of LOS flux blocking from geometrical optics remains valid, corrections from wave effects substantially alter the observed flux variation.
In particular, the mask induces a `negative' point spread function (PSF) on the background source; thus, the light-blocking intuition should be understood in relation to the PSF-smeared mask.
\end{enumerate}
It remains possible to infer the angular size of stars in the wave optics regime by measuring light curves during occultations.
First, we will derive the negative-PSF induced by the mask on a point source of light of wavelength $\lambda$.
%This can be computed using the Fresnel diffraction integral.
If the angular size of the mask is much larger than the star, the geometry of the mask can be approximated as a straight edge \cite{occultation-straight-edge-derive}.
With this negative aperture (negative in the sense that it blocks light instead of allowing light to pass), the negative-PSF is obtained from the Fresnel diffraction integral\footnote{Note that the near-field approximation $D_{\rm OM}/D_{\rm MS} \ll 1$ has been used in the derivation of this equation. Consequently, the mask-to-source distance $D_{\rm MS}$ does not appear, making it impossible to recover the geometrical optics limit from this equation alone.},
\begin{equation}
\begin{split} \label{eq: fresnel-diffraction-integral}
    I(x) &= I_0 \left( \left|\int_{-\infty}^{x}\cos(\pi r^2/D_{\rm OM}\lambda)  \, dr\right|^2 + \left|\int_{-\infty}^{x}\sin(\pi r^2/D_{\rm OM}\lambda)  \, dr\right|^2 \right) \\
    &\equiv I_0 \left( \left(\mathcal{C}(x/D_F)+\frac{1}{2}\right)^2 + \left(\mathcal{S}(x/D_F)+\frac{1}{2}\right)^2\right),
\end{split}
\end{equation}
where $x$ denotes the location of the point source relative to the geometrical edge of the mask.
The Fresnel cosine integral\footnote{Multiple definitions exist of the Fresnel cosine/sine integrals. We adopt conventions used by \textsc{scipy}.special.fresnel().} $\mathcal{C}(x)$ and the Fresnel sine integral $\mathcal{S}(x)$ produce a damped oscillatory pattern with characteristic scale $D_F \equiv \sqrt{\lambda D_{\rm OM}}/2$.
In astronomical applications, we are interested in the \textbf{angular} Fresnel scale perceived by the observer, %$\theta_F \equiv D_F/D_{\rm OM} = 0.5\sqrt{\lambda/D_{\rm OM}}$.The typical angular Fresnel scale is:
\begin{equation} \label{eq: fresnel-scale}
    \theta_F \equiv \frac{D_F}{D_{\rm OM}} = 3.6 \left(\frac{\lambda}{500 \mathrm{ \,nm}}\right)^{1/2}\left(\frac{D_{\rm OM}}{4\times 10^5 \mathrm{\,km}}\right)^{-1/2}  \mathrm{milliarcseconds},
\end{equation}
where the characteristic value $4\times 10^5$\,km is the average Earth--Moon distance.
Since the typical scale of $\theta_F$ is below the diffraction limit $\theta_{\rm res}$ of current telescopes, the occultation pattern cannot be spatially resolved.

Consider a star with angular diameter $\theta_*$ that is even smaller than the Fresnel scale, $\theta_* < \theta_F < \theta_{\rm res}$. 
A resolution-limited observer can measure only the total flux from the star.
However, the total flux is the sum\footnote{Here, we implicitly assume that the wavefronts emitted at $\theta'$ and $\theta'+d\theta'$ on the star are uncorrelated, allowing the intensities to be summed directly without accounting for wave optics effects. This holds because the relevant scale of $d\theta'$ corresponds to the optical path difference defined by $\lambda$ rather than the Fresnel scale.} of diffraction patterns from different regions of the stellar disc. %, at different distances $\lbrace x \rbrace$ from the mask. 
Integrating equation~\eqref{eq: fresnel-diffraction-integral} across the stellar disc at a moment when the angular separation between the centre of the stellar disc and the edge of the mask is $\theta$, the total flux is
\begin{equation} \label{eq: finite-size-flux} 
\begin{split}
    \mathcal{F}(\theta; \theta_*) &\equiv \int_{\theta-\theta_*/2}^{\theta+\theta_*/2} I(\theta') d\theta' \\
    &\propto \int_{\theta-\theta_*/2}^{\theta+\theta_*/2} \mathcal{G}(\theta') \left( \left(\mathcal{C}(\theta'/\theta_F) + \frac{1}{2}\right)^2  +  \left(\mathcal{S}(\theta'/\theta_F) + \frac{1}{2}\right)^2 \right)d\theta'.
\end{split}
\end{equation}
The leading `geometrical' factor $\mathcal{G}(\theta')$ is determined by the surface brightness distribution across the star; two specific cases are
\begin{equation}
     \mathcal{G}(\theta') = \begin{cases}
         2\sqrt{(\theta_*/2)^2-\theta'^2} & {\text{uniform circular disc, with $\theta'<\pm \theta_*/2$}}, \\
         \mathrm{constant} & {\text{uniform box of width $\theta_*$}}.
     \end{cases}
\end{equation}
In general, we shall solve integral~\eqref{eq: finite-size-flux} numerically. However, to understand the effect of a source with finite diameter on the diffraction pattern, we can use the WKB method (and the uniform box model) to approximate the perturbation from a genuine point source
\begin{equation} \label{eq: amplitude-suppress-approx}
   \Gamma = \frac{\mathcal{F}(\theta; \theta_*)-\mathcal{F}(\theta; \theta_*=0)}{I_0} \approx \frac{\sqrt{2}\theta_F}{\pi \theta} \left[ \underbrace{\left(1 - \text{sinc}\left(\frac{\pi\theta \theta_*}{2\theta_F}\right)\right) \cos\Phi(\theta)}_{\text{Amplitude Suppression}} + \underbrace{\frac{\pi \theta_*^2}{8} \text{sinc}\left(\frac{\pi\theta \theta_*}{2\theta_F}\right) \sin\Phi(\theta)}_{\text{Phase Misalignment}} \right],
\end{equation}
where the phase factor $\Phi(\theta) = (\frac{\pi}{2} (\theta/\theta_F)^2) + \pi/4$.
The net effect of a star with finite angular size $\theta_*$ is to suppress the intensity of Fresnel fringes (particularly after the first fringe), and to slightly shift their phases. The effective value of $\theta_F$ does not change.
It is measurements of the fringe intensity that will provide the most constraining power on $\theta_*$. 
Importantly, the fringes are not uniformly spaced: the separation between turning points in $\mathcal{F}(\theta)$ decreases as $\propto 1/\theta$. To measure the amplitude of second, third and fourth fringes, a detector must resolve the diffraction pattern {\it better} than just the first fringe at $\theta=2\pi/\theta_F$. 
In practice, even higher order fringes are further truncated by the finite bandwidth of photons, as $\theta_F \propto \lambda$. 
In a broadband photon sensor, the higher fringes from different wavelengths would lose phase coherence, thus destroying the fringe pattern. 
In the numerical results that follow, we fix the bandwidth to be $\Delta \lambda = \pm 50 $ nm to avoid over-fitting to unphysically high order fringes.

\subsection{Observable consequences}

Occultation imaging relies on a foreground mask that moves relative to a distant source at a finite angular speed. 
If the angular offset between the mask and the source $\theta(t)$ changes over time, so does the received flux (Eq.~\ref{eq: finite-size-flux}).
Although the Fresnel scale induced by the mask is well below the telescope's spatial resolution limit, {\it temporal} variations in the light curve remain observable.

The {period} of oscillations in a light curve provides a measurement of $\theta_F$, allowing us to infer the negative PSF and thus $\mathcal{F}(\theta(t); \theta_* = 0)$.
The observed {amplitude} of oscillations in the light curve then measure the star's angular size.
The measurement of the timescale and the fringe amplitude are thus largely independent.
This effect is observable only with fine time resolution, as the oscillations in flux are prominent only when the mask aligns within certain multiples $\theta_F$ of the source, beyond which the wave-optical coherence is lost. 
Furthermore, the spacing between higher order fringes is increasingly compact, which poses a stringent requirement for temporally resolving the fringes.
The whole occultation event lasts a duration of order $T_F$, the time required for the mask to move across a Fresnel scale. 

\begin{figure}[!t]
    \centering
    \includegraphics[width=0.9\linewidth]{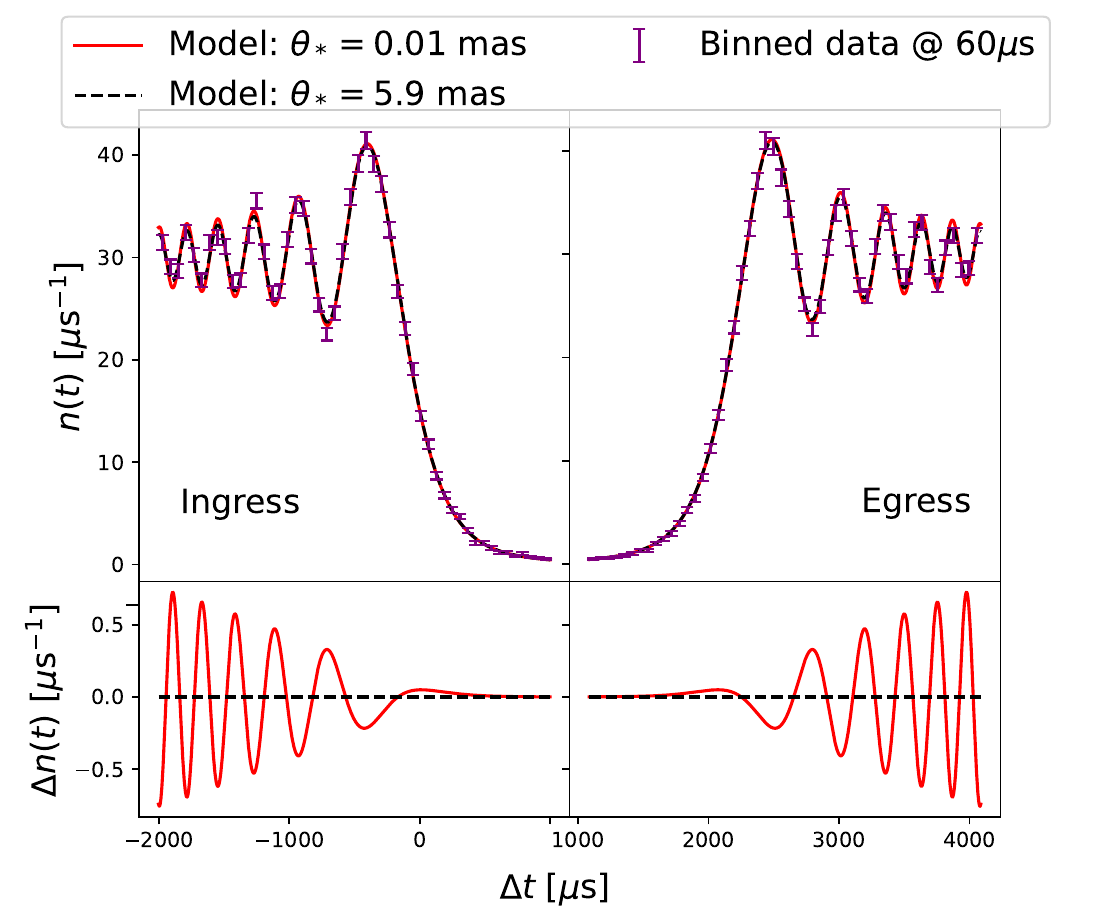}
    \caption{Top panel: Light curves of a distant object of diameter $\theta_* = 5.9$~mlliarcseconds (dashed line), or an effectively pointlike object (solid line), as they enter then emerge from occultation by a GPS satellite. Both objects normally produce $30$ photons at $\lambda=500$nm every $1 \mu$s.
    Error bars show a random realisation of data, binned every $60 \mu $s. 
    Bottom panel: The two light curves look similar up to the first Fresnel fringe, then their amplitudes differ (see eqn~\ref{eq: amplitude-suppress-approx}). The amplitude of $\Delta n(t)$ continues to increase beyond the limits of this figure, before returning to zero.}
    \label{fig: lightcurve-example}
\end{figure}

An example light curve in which a distant star with angular diameter $\theta_* = 5.9$ mas is occulted by a GPS satellite flyby is shown in Fig.~\ref{fig: lightcurve-example} .
For comparison, we also overlay the light curve of an extremely small object ($\theta_* = 0.01$ mas), which is effectively a point source.
The difference between the two light-curve models is small.
The maximum difference between the two light curves occurs at their stationary points; the uncertainty in measuring the angular size of stars will be determined by the precision with which the amplitude of the Fresnel fringes can be measured, in these moments that last only a fraction of $T_F$.
In this case, the timescale around the points of maximal difference is approximately $10 \mu$s.
A photon-detecting device with microsecond-level resolution will be essential to detect this subtle signal.

The typical angular velocity $\omega$ of the mask relative to the distant star can be estimated via Kelper's law.
Alternatively, if the masks are artificial objects, we might track their instantaneous velocity using their on-flight tracking system.
The instantaneous angular velocity vector might not align with the normal of the occultating edge of the mask.
Denote this bisection angle as $\phi$.
Consequently, the projected angular speed of the mask would be moving at a discounted factor of $\cos(\phi)$, which relates the angular offset of the mask and observation time as: 
$\theta(t) = \omega t \cos\phi$.
This simplistic estimation shows that it is possible to scan across the Fresnel fringes in the time domain.  
As the angular size of the fringe is fixed by the geometry (Eq.~\ref{eq: fresnel-scale}), the unknown factor $\omega\cos\phi$ can be determined empirically from the data.

\subsection{Definition of effective angular resolution} \label{sect: resolution}

The resolution of standard telescope imaging can be defined by the Abbe criterion or the Rayleigh criterion, 
\begin{equation}
    \theta_{\rm Rayleigh} \equiv 1.22 \frac{\lambda}{A},
\label{eq: rayleigh}
\end{equation}
where $A$ is the diameter of the telescope's circular aperture and $\lambda$ is the observing wavelength.
The numerical factor of $1.22$ roughly corresponds to the full-width half-maximum of a typical PSF.
From a more statistical perspective, the choice of $1.22$ is an SNR-based argument.
If two point sources are separated by a distance smaller than $\theta_{\rm Rayleigh}$, the two PSFs merge.
Consequently, there is no longer a sufficient SNR at which it becomes possible to assert that there are two separate sources.

\begin{figure}[t]
    \centering
    \includegraphics[width=0.9\linewidth]{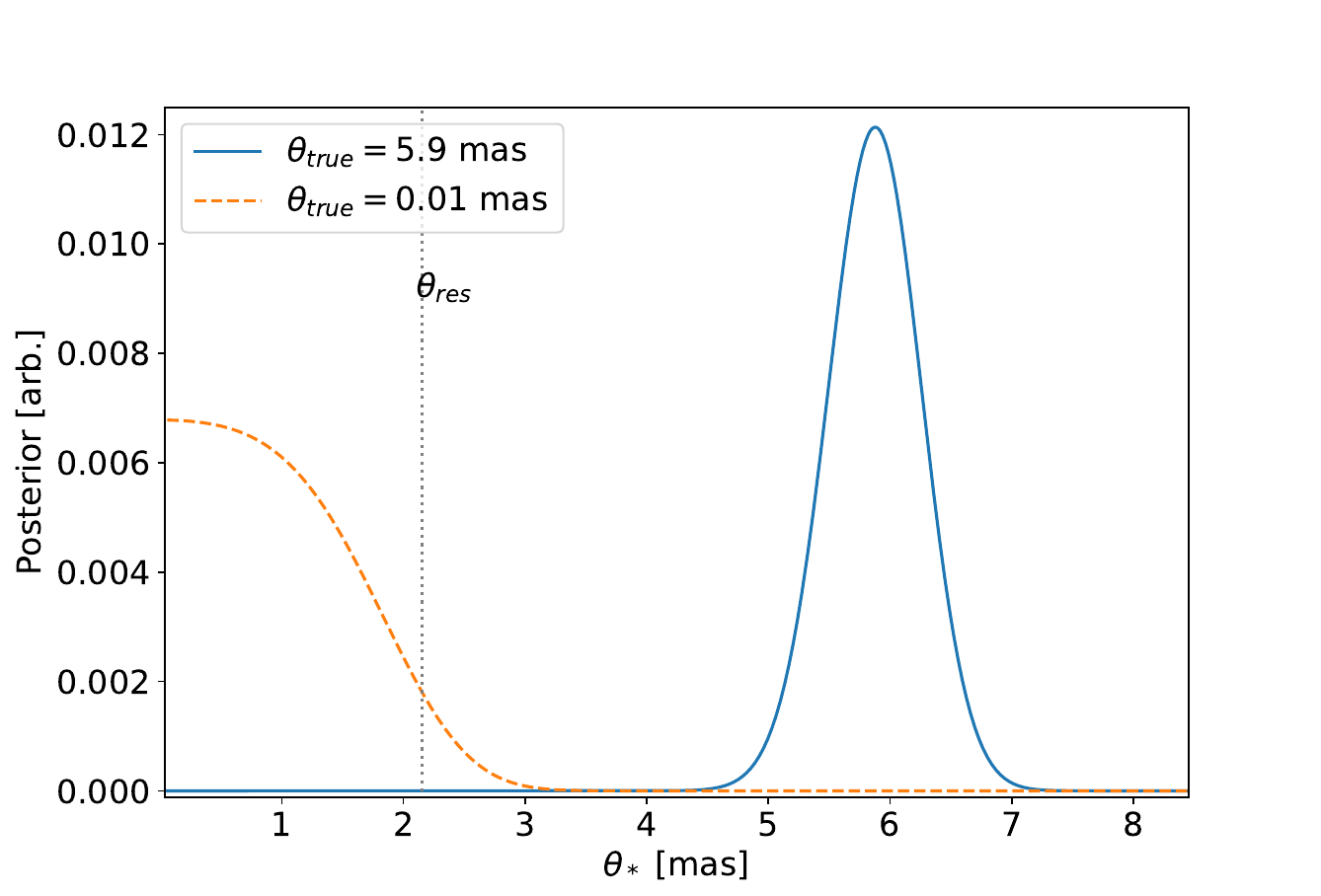}
    \caption{The unnormalised posterior probability of the inferred size of a star that is genuinely pointlike (red, $\theta_{\rm true} = 0.01$ mas) and a resolvable object (blue, $\theta_{\rm true} = 5.9$ mas). This calculation assumes an observing setup like that in Fig.~\ref{fig: lightcurve-example}, and uses the data analysis pipeline described in Section~\ref{sect: stats}. Crucially, we define the resolution limit of occultation imaging, $\theta_{\rm res}$, by the $95\%$ upper bound on the inferred angular size of a pointlike source (dotted vertical line). %Note that the posterior reported here is not normalised. 
    }
    \label{fig: example-posterior}
\end{figure}

Using a SNR-based argument in the same spirit, we define an effective angular resolution for occultation imaging, $\theta_{\rm res}$. We adopt the $95$th percentile of the posterior in the inferred size, $\theta_{*}$, of a genuine point source from time series data $\lbrace\mathcal{F}(t_i)\rbrace$:
\begin{equation}
    0.95=\int_0^{\theta_{\rm res}} \mathrm{Pr}(  \theta_* | \lbrace \mathcal{F}(t_i)\rbrace, \Delta T, \theta_F) \, d\theta_*.
\end{equation}
As such, it is $0.95/(1-0.95)\approx 19$ times more likely that the source is smaller than $\theta_{\rm res}$ than larger.
%This definition is indicated by the dashed line in Fig.~\ref{fig: example-posterior}.
%An effective point source with an angular size of $\theta_{\rm true} = 0.01$ mas is injected into the simulation pipeline, and the posterior is evaluated for this setup.
The resulting posterior is one-sided, favouring the minimum possible value for $\theta_*$ allowed by the prior (Fig.~\ref{fig: example-posterior}).
%The vertical dashed line indicates $\theta_{\rm res}$, beyond which the model is strongly disfavoured by the data.

From a purely optical perspective, the resolution of occultation imaging does not depend on the wavelength of observations $\lambda$ (unlike standard diffraction-limited imaging, eq.~\ref{eq: rayleigh}). This statement is not trivial. At long wavelengths, an increase in the Fresnel scale $\theta_F \propto \lambda^{1/2}$ reduces the light curve damping effect from the finite stellar size, resulting in lower angular resolution. However, for a source with a spectral energy distribution that is flat 
% in $d\lambda/\lambda$, \rjm{[is that correct?]}
(in the sense that $f_\lambda(\lambda) \propto \lambda$),
this is counterbalanced by the conversion from apparent magnitude (energy flux) to the photon count rate. An increased number of long wavelength photons, each carrying less energy, results in a higher Poissonian SNR $\sim \lambda^{1/2}$.
Consequently, this effect balances the degradation in the Fresnel scale, hence removing the wavelength dependence.
In practice, stars do not have a flat SED, and the quantum efficiency of the photon sensor is usually a highly sensitive function of wavelength. While these effects can be modelled in actual observations, to maintain the generality of this work, we shall simply assume a fixed observing wavelength $\lambda_{500} = 500$ nm, using the subscript as compact notation to remind the reader of this assumption. 

\subsection{Available foreground masks}
\begin{table}
\centering
\begin{tabular}{| c | c | c | c | c || c | c |} 
 \hline
 Mask & $\omega$ [mas $\mu$s$^{-1}$]& $\theta_F$ [mas]&  $T_F$ [$\mu$s] & $D_{\rm OM}$ [km] & $\theta_{\rm geo}$ [mas] & $N_{\rm mask}$ \\ 
 \hline\hline
 Starlink satellites (LEO) & $2.3 \times 10^{-1}$ & 98 &  $4.3\times 10^2$ & $550$ & $3.8\times 10^3$& $10^4$\\ % this is the sidereal speed, achieved by using an active star tracker to cancel the Earth's rotation
 % Starlinks & $2.7 \times 10^{0}$ & $98$ &  $36$ \\ % this is the apparent speed without taking into account of Earth's rotation
 \hline
  % MEO satellite & $1.3 \times 10^{-2}$ & 13 &  400 \\
 GPS-like satellites (MEO) & $3.5 \times 10^{-2}$ & $16$ & $4.6 \times10^2$ & $2\times 10^4$ & $1.0 \times 10^2$ & $10^{2~\dagger}$  \\
 \hline
 % Moon & $5.5\times 10^{-4}$ &3.6 & $6.5\times 10^3$ \\
 The Moon & $3\times 10^{-4}$ &3.6 & $1.2\times 10^4$ & $3.6\times 10^5$& $1.9\times 10^6$ & 1\\
  % Moon & $1.5\times 10^{-2}$ &3.6 & $2.4\times 10^2$ \\
 \hline
 % Asteroids & & & \\
 % \hline
 \multicolumn{7}{c}{$^{\dagger}$\footnotesize{The number reported here includes satellites that share similar orbits with the GPS satellites.}}
\end{tabular}
\caption{Numerical parameters adopted to characterise different masks. Columns show the approximate number of objects in each class $N_{\rm mask}$, their typical geometric angular size $\theta_{\rm geo}$ and distance from Earth $D_{\rm OM}$, their angular speed $\omega$ and Fresnel crossing time $T_F$, and their Fresnel scale $\theta_F$, calculated assuming the observation wavelength is $\lambda = 500$\,nm. The angular speed of objects is reported in the frame of an observer on the surface of the Earth, assuming the Earth's rotation is cancelled by an active star tracking system. As such, the resulting angular speeds are effectively with respect to a distant background star. %The two rightmost columns are only relevant for event rate calculation but do not contribute to the wave-optical light curve calculation. 
}
\label{tab: occultation-mask-param}
% \end{minipage}
\end{table}

Several classes of object move in the night sky and can potentially be used as foreground masks (Table~\ref{tab: occultation-mask-param}). The rate at which they produce occultation events depends upon the fraction of sky covered by their tracks --- their projected angular diameter $\theta_{\rm geo}$ multiplied by their angular speed $\omega$ --- and the number density of sufficiently bright stars.
With most being artificial or natural satellites bound by the Earth's gravity, their average angular speed depends only on their distance $D_{\rm OM}$, via Kepler's third law, $\omega \propto (R_\oplus+D_{\rm OM})^{-3/2}$, where $R_\oplus$ is the radius of the Earth.
In principle, asteroids could also be used; however, as their orbital characteristics vary on a case-by-case basis, we shall not consider them here.
Case studies of asteroid occultation can be found in, for example, \cite{asteroid-occultation-2024-sCMOS,asteroid-occultation-obs-2009-ccd}.

During the short timescale $T_F$ of most occultation events, flux measurement precision will be limited by Poissonian photon-counting shot noise; thus, the SNR for constraining the flux scales as $\sim \sqrt{T_F}$.
Consequently, brighter stars allow for more accurate flux measurements, and in principle, their angular diameters can be measured with the highest precision.
The only exception is that light from the Moon, even during its darkest phase, typically dominates the noise budget rather than the source star's Poissonian noise.
As a result, measuring stellar diameters via occultation by MEO satellites is not necessarily less accurate than using the Moon.

Notably, the occultation timescale $T_F$ for LEO satellites is the same order of magnitude as that for MEO satellites, leading to similar Poissonian photon-counting noise.
This results from their lower orbit; the larger angular Fresnel scale $\theta_F \propto D_{\rm OM}^{-1/2}$ is compensated by a faster angular speed, in accordance with Kepler's third law.
However LEO satellites are less sensitive to $\theta_*$ overall, because their lower ratio $\theta_* / \theta_F$ leads to a smaller change in light-curve amplitude ($\Gamma$ in Eq.~\ref{eq: amplitude-suppress-approx}).
Nonetheless, the Fresnel scale for Starlink satellites is larger than the resolution of a Hubble-like telescope.
With sufficiently fast imaging to capture the event, the negative-PSF of a Starlink occultation should be {\it spatially} resolvable as well.

\section{Simulated data} \label{sect: stats}
\subsection{Generation of mock observations}

To create mock photometric observations of an occultation, we assume the angular speed of the occulting mask is known from external constraints with high precision \cite{starlink_impact_halferty}.
The offset of the occulting mask from the centre of the source star $\theta$ can then be parametrised by time: $\theta = \omega t $. 
If the time resolution of the detector is $\Delta T < T_F$, and there is no dead time between exposures, a series of observations at time steps $\lbrace t_i = t_0 + i\Delta T \rbrace$ produces a light curve $\mathcal{F}_i\equiv \mathcal{F}(\theta(t_i); \theta_*)$ that is a temporal scan through the negative PSF (Eq.~\ref{eq: finite-size-flux}).
This can be translated into observed photon counts $\lbrace n(t_i)\rbrace=\lbrace\mathcal{F}_i\Delta T\rbrace$.

\begin{figure}[t]
    \centering
    \includegraphics[width=0.9\linewidth]{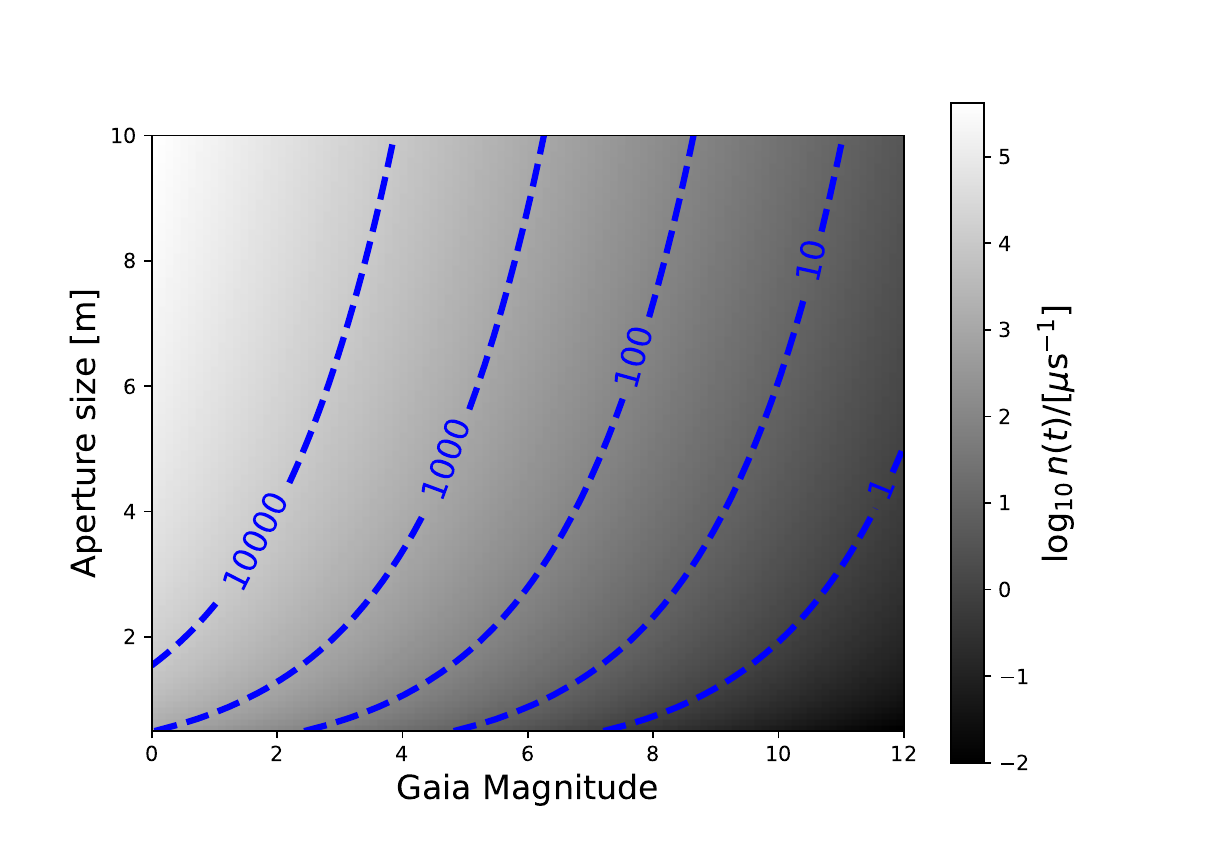}
    \caption{The mean number of photons $n_0$ received from a star per microsecond (before or after an occultation), as a function of the star's Gaia G-band magnitude \cite{gaia-dr2} and the telescope's effective aperture. This assumes telescope throughput and detector quantum efficiency of $30\%$ at wavelength $\lambda=500$\,nm, as is typical for ultra-fast photon detectors with sub-microsecond temporal resolution.
    \label{fig: photon-count-to-brightness}}
\end{figure}

%To evaluate the attainable angular resolution, we set up simulated observations of occultations using different classes of foreground masks.
The amplitude of the light curve --- the number of detected photons --- scales proportionally with the star's brightness and is, in general, a function of its spectral density distribution, the observing wavelength and bandpass, the telescope's light-collecting area, and the detector's quantum efficiency.
To simplify the discussion, we vary the {\it effective} parameter: the mean number of photons $n_0$ received per microsecond before or after the transit.
%This avoids the complication of additional Poissonian noise originating from the masks themselves, whilst still maintaining a reasonable SNR. 
%We repeated the same analysis assuming various nominal values of photon flux, at $n(t) = \lbrace30,100,1000\rbrace \, \mu$s$^{-1}$. 
Mean photon flux values can be translated into a function of the apparent magnitude of the background stars and the effective aperture size of the telescope (Fig.~\ref{fig: photon-count-to-brightness}).

To add observational noise, we note that any observation with microsecond exposure times will likely be in the photon-starved regime. With input values of $\Delta T$, $\theta_F$ and $\theta_*$, we draw random samples of the `observed' number of photons $\lbrace n(t_i)\rbrace$ from a Poisson distribution:
\begin{equation} \label{eq: likelihood}
    \mathrm{Pr}( \lbrace n(t_i)\rbrace |\Delta T, \theta_F, \theta_*, n_0) \propto \prod_{\lbrace t_i\rbrace} (\mathcal{F}_i\Delta T)^{n(t_i)}\exp(-\mathcal{F}_i\Delta T).
\end{equation}
We assume that other sources of photons are negligible, including detector dark current\footnote{On-sky performance \cite{lau2022development} of the SPINA experiment \cite{lau2023initial} achieved dark count rate of $\sim$$1$ photon per millisecond \cite{spina-noise-model}, which is sub-dominant and effectively zero in our simulations.} and reflected Earthshine from artificial satellites\footnote{This is appropriate so long as occultations are observed when the Sun is well below the horizon \cite{starlink_impact_mroz,starlink_impact_mendoza}.}. 
In the special case of lunar occultation, photons reflected from the Moon become an additional, dominant source of statistical fluctuations.
Assuming the flux from the Moon $\mathcal{F}_{\rm Moon}$ is constant over timescale $T_F$, %\rjm{How good is this approximation? Relative to the few-photon statistics of the background star? The Moon is moving, and we'd be observing an aperture into which the edge is moving} 
the expectation value for the number of observed photons would be $n(t_i) = \mathcal{F}_{\rm Moon} + \mathcal{F}_i$, and the likelihood of observing a time series of photon counts would be
\begin{equation}
    \mathrm{Pr}( \lbrace n(t_i)\rbrace |\Delta T, \theta_F, \theta_*, n_0, \mathcal{F}_{\rm Moon}) \propto \prod_{\lbrace t_i\rbrace} ((\mathcal{F}_{\rm Moon}+\mathcal{F}_i)\Delta T)^{n(t_i)}\exp(-(\mathcal{F}_{\rm Moon}+\mathcal{F}_i)\Delta T),
\end{equation}
which increases the variance of the distribution, thereby degrading its constraining power.
Even a new Moon has apparent magnitude $V_{\rm Moon} = 0.15$ \cite{darkmoon-flux-obs}, overwhelming the photons from a $V=8$ star by a factor $\sim$$10^3$. In our simulations, we assume that it is possible to physically mask or digitally subtract the flux contributed from $93\%$ of its area, leaving apparent magnitude $V_{\rm Moon-masked} = 2$. We denote this mask as the 7\% new Moon.

\subsection{Analysis of mock observations}
When fitting the shot noise-limited data, we propagate uncertainty using a Bayesian model, as the classical $\chi^2$ statistic is not a good approximation.
Indeed, Bayes' Theorem provides the posterior probability for the diameter of the background star, $\theta_*$,
\begin{equation}\label{eq: posterior}
    \mathrm{Pr}(  \theta_* | \lbrace n(t_i)\rbrace, \Delta T, \theta_F, \mathcal{F}_{\rm moon}, n_0) \propto \mathrm{Pr}( \lbrace n(t_i)\rbrace |\Delta t, \theta_F, \theta_*, \mathcal{F}_{\rm moon},n_0) \cdot \pi(\theta_*),
\end{equation}
where $\pi(\theta_*)$ is a prior, which we assume to be uniform between zero and $\theta_F$. Parameters $\mathcal{F}_{\rm moon}$ and $n_0$ can be measured to arbitrary precision by taking long exposures of the stars before/after the occultation. 
As such, we treat them as observables instead of unknown parameters to be marginalised over.

Since measurements of $\theta_*$ are essentially one-tailed upper limites, we record the 95th percentile highest density interval (HDI) of this posterior (consistently with our definition of resolution in Section~\ref{sect: resolution}).

\begin{figure}[!t]
    \centering
    \includegraphics[width=0.9\linewidth]{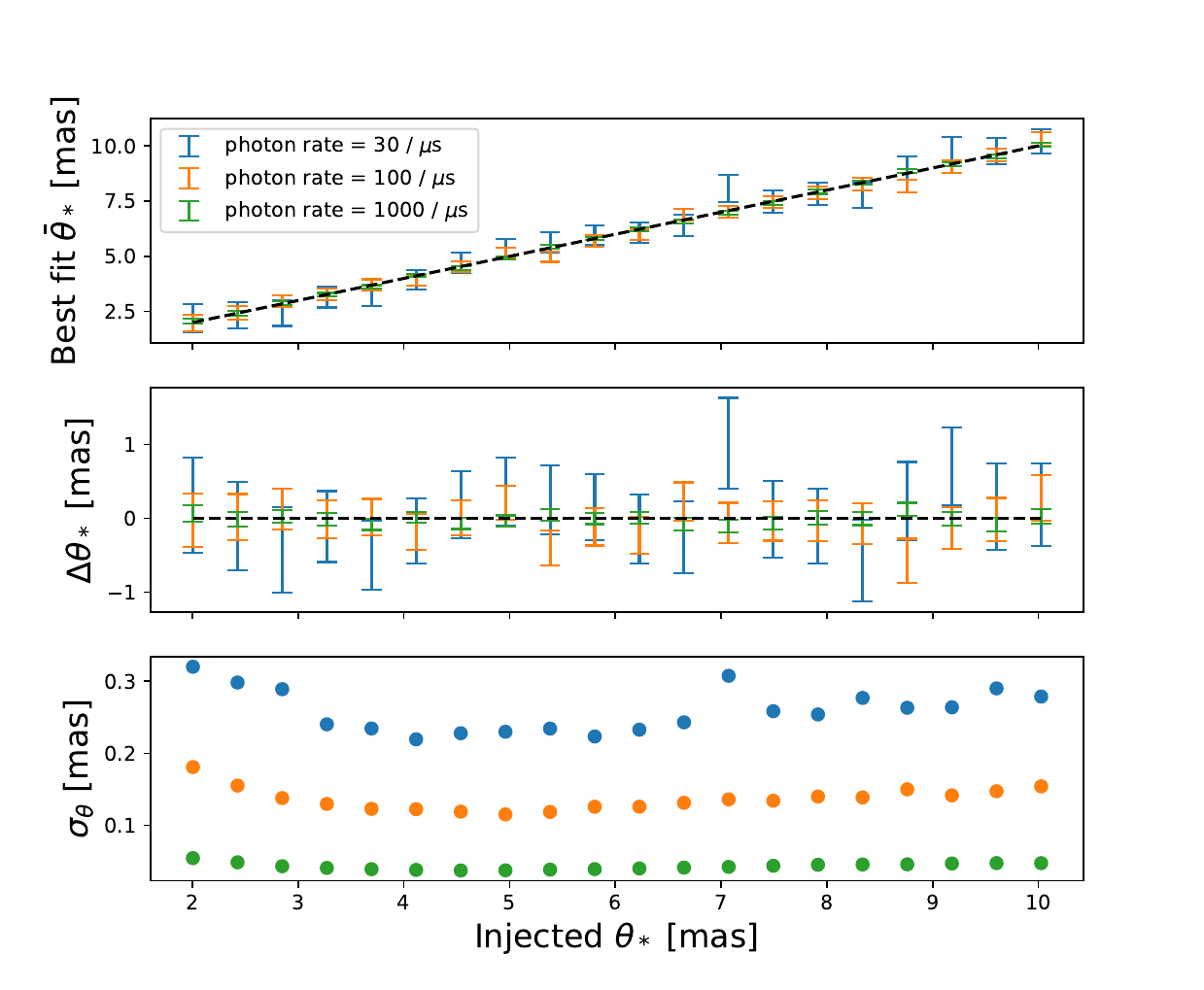}
    \caption{Validation that the proposed Bayesian inference pipeline can measure stellar diameters $\theta_*$ without bias, from simulated, noisy light curves of occultation by satellites in Medium-Earth Orbit (MEO).
    The middle panel shows bias from the injected ground truth, consistent with zero.
    The lower panel shows the $1\sigma$ measurement  uncertainty for three different observing scenarios that encapsulate different stellar brightness, telescope aperture, etc.}
    \label{fig: injected-vs-inferred-size-MEO}
\end{figure}

\subsection{Validation of data analysis pipeline}

We simulate light curves for MEO satellite occultations of a pointlike star to set up a baseline, plus a population of stars with diameters ranging from $2$ to $10$\,milliarcseconds (Fig.~\ref{fig: lightcurve-example} shows two examples). 
These stars are unrealistically large, but the range is broad enough to ensure that the inferred angular size has well-defined upper and lower bounds, while remaining smaller than the $\theta_{\rm Rayleigh}$$\sim$$10^1$\,milliarcsecond optical diffraction limit. 
From noisy, simulated photon time streams, we infer the Bayesian posterior distribution (Eq.~\ref{eq: posterior}) for the angular size of the star, $\theta_*$.
Fig.~\ref{fig: example-posterior} shows two examples.

To check that our analysis pipeline can recover the injected parameters in a high signal-to-noise regime, we simulate mock observations of resolvable stars. We then run the inference pipeline for each injected value of stellar diameter $\theta_*$ and flux $n_0=\lbrace30,100,1000\rbrace \, \mu$s$^{-1}$. The maximum a-posteriori (i.e.\ the best-fit) value $\bar{\theta}_*$ is consistent with the injected value (Fig.~\ref{fig: injected-vs-inferred-size-MEO}). 
We find that these posteriors, far from the edges of the prior, are well approximated by a Gaussian. 
Since measurements for these resolvable sources are two-tailed, we report measurement uncertainty in terms of Gaussian $\sigma_\theta$.
We find inference bias $\Delta\theta_* \equiv \bar{\theta}_* - \theta_*$ consistent with zero, validating the method in this regime.
%\rjm{The behaviour of the curves in Fig.~\ref{fig: injected-vs-inferred-size-MEO} looks pretty predictable... adding two terms in quadrature? Could you fit a function to it? Could you even guess the shape of the function from first principles? It would be more use to other people if you have a handy analytical approximation.}

\section{Results}
\label{sect: results}

\subsection{Attainable resolution limit}

The estimation in Section.~\ref{sect: formalism} demonstrates that the choice of different foreground masks could significantly change the sensitivity, and thus the resolution obtained from occultation. 
Henceforth, we will factor the photon rate into the telescope aperture size and the apparent magnitude of the background star, as these are conventionally used in survey planning. 
We use the aforementioned pipeline to estimate the resolution limit as a function of telescope aperture size for Low-Earth Orbit (LEO) satellites, Global Positioning System (GPS) satellites, and the New Moon 
\footnote{The New Moon is defined so that the apparent magnitude of the moon is $m_G=2$, which corresponds to $\sim 7\%$ of the peak brightness of the full moon at its monthly brightest phase.}
as the foreground occulting mask.
We selected characteristic values for the background stars of $m_G = \lbrace 6, 10\rbrace$, which, when combined with the aperture size and optical throughput, fixed the photon flux as shown in Fig.~\ref{fig: photon-count-to-brightness}. 

\begin{figure}[t]
    \centering
    \includegraphics[width=0.95\linewidth]{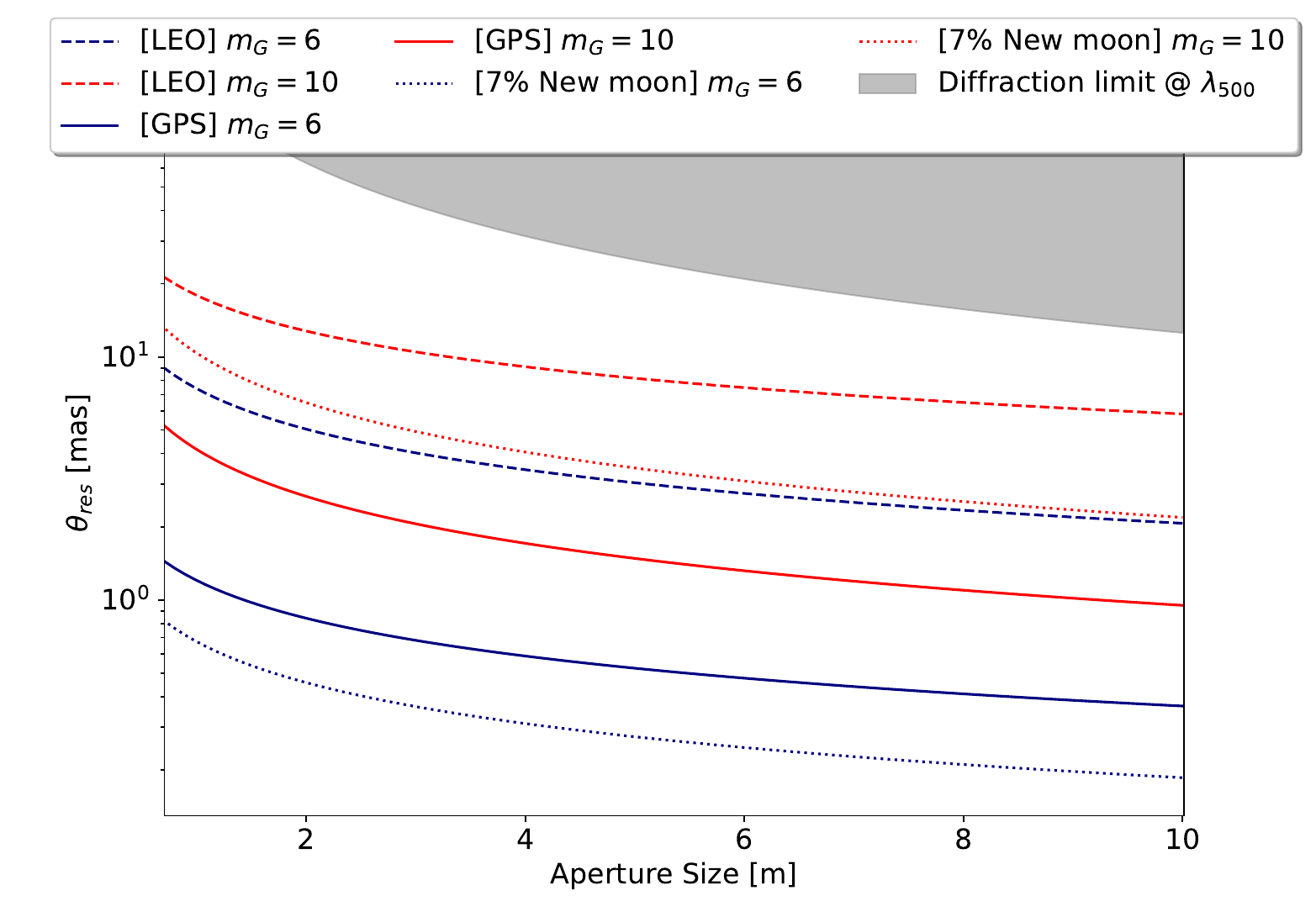}
    \caption{The attainable resolution limit $\theta_{\rm res}$ with the occultation method on stars of different magnitudes, compared to the optical diffraction limit. Here, the diffraction limit of optical telescopes is defined at the observing wavelength $\lambda_{500} \equiv 500$ nm. The occultation-based resolution only weakly scales with the photon wavelength, by the fact that the Poissonian noise of shorter wavelength photons is stronger at the same flux as each of the shorter wavelength photons carry less energy.}
    \label{fig: res-lim-appertures}
\end{figure}

The results are shown in Fig.~\ref{fig: res-lim-appertures}. For reference, we display the optical diffraction limit at $\lambda_{500} = 500$ nm as the shaded area, below which the system operates in the super-resolution regime. 
The LEO satellites, indicated by dashed lines, show only a mild improvement over the diffraction limit.
This is due to their proximity to the observer (on the Earth's surface), resulting in a larger Fresnel scale $\theta_F$.
Satellites in Medium Earth Orbit (MEO), primarily GPS and similar navigation satellites (solid lines), can easily achieve a resolution of $\theta_{\rm res} < 10$ mas.
For bright stars observed with a large telescope, the resulting resolution can even reach $\theta_{\rm res} < 1$ mas. 
It is also worth noting that the improvement in $\theta_{\rm res}$ grows more slowly with increasing aperture size relative to the diffraction limit $\theta_{\rm diff} \propto 1/D$.
An extremely large telescope equipped with adaptive optics could attain better angular resolution through diffraction-limited imaging than through occultation imaging.
That being said, the engineering of telescopes much larger than $10$ m requires precise and ultra-fast control to compensate for atmospheric instability, and the target performance of the forthcoming European Southern Observatory Extremely Large Telescope is yet to be verified. 

Lastly, traditional lunar occultation (dash-dotted lines) outperforms satellite-based occultation, as the associated Fresnel scale $\theta_F$ is substantially smaller and the timescale of the light-curve variation is much longer. 
With a reasonable mask applied, the angular resolution can easily reach $\theta_{\rm res} \sim 10^{-1}$ mas. 
However, the photon flux from the Moon dominates the Poissonian noise budget, resulting in a rapid decline in resolution as the background star becomes dimmer. 
Evidently, GPS satellite occultation outperforms lunar occultation when the background star is dimmer than $m_G \gtrsim 10$. 

\subsection{Occultation Cross Section and Rate}
\begin{figure}[!t]
    \centering
    \includegraphics[width=0.95\linewidth]{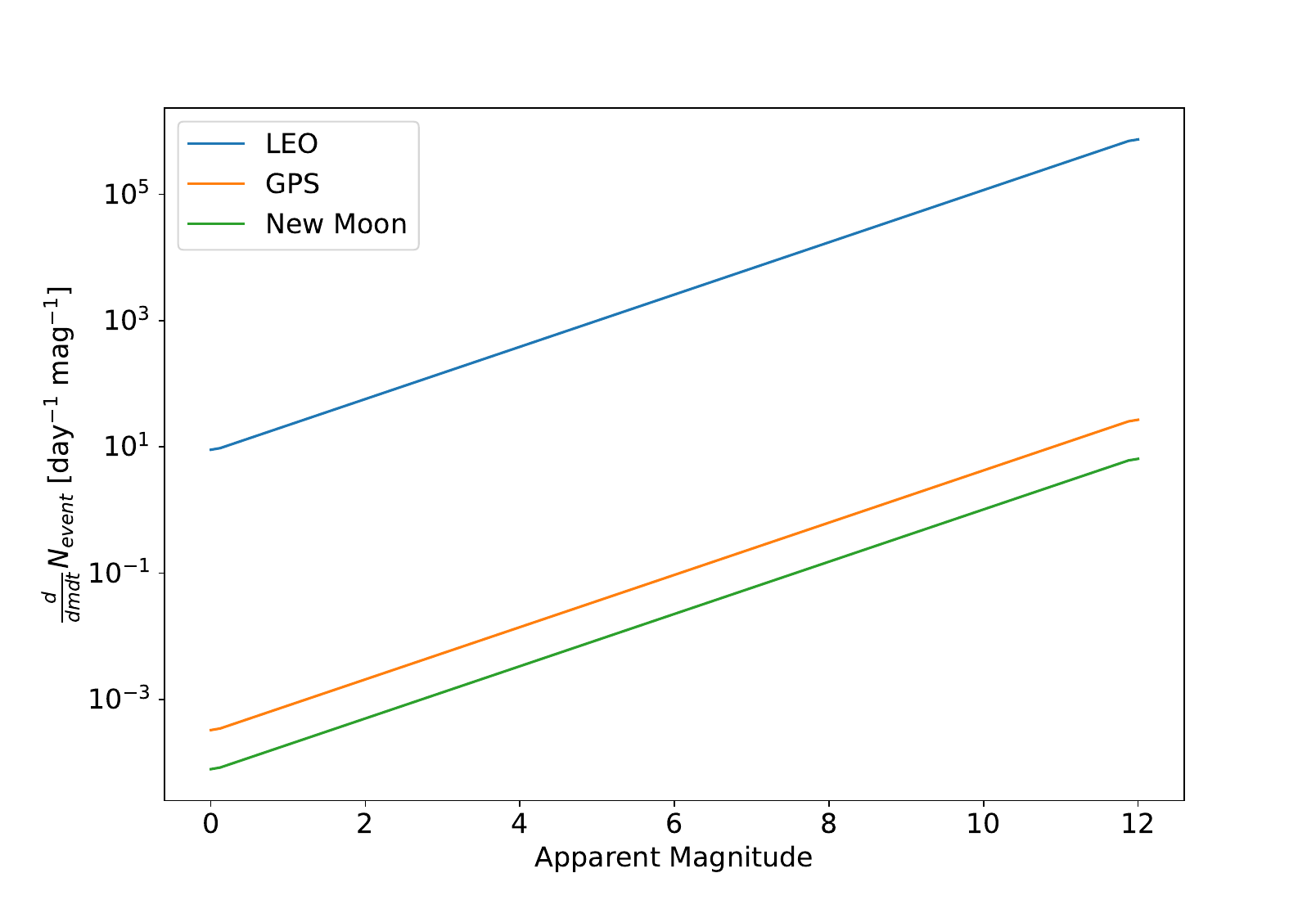}   
    \caption{The occultation event rate of different foreground mask choices as a function of apparent magnitude of background stars. The apparent magnitude of stars is critical to determine how much Poissonian signal-to-noise ratio, thus the reconstruction uncertainty in stellar angular size.
    }
    \label{fig: event-rate}
\end{figure}

\begin{figure}[!t]
    \centering
    \includegraphics[width=0.95\linewidth]{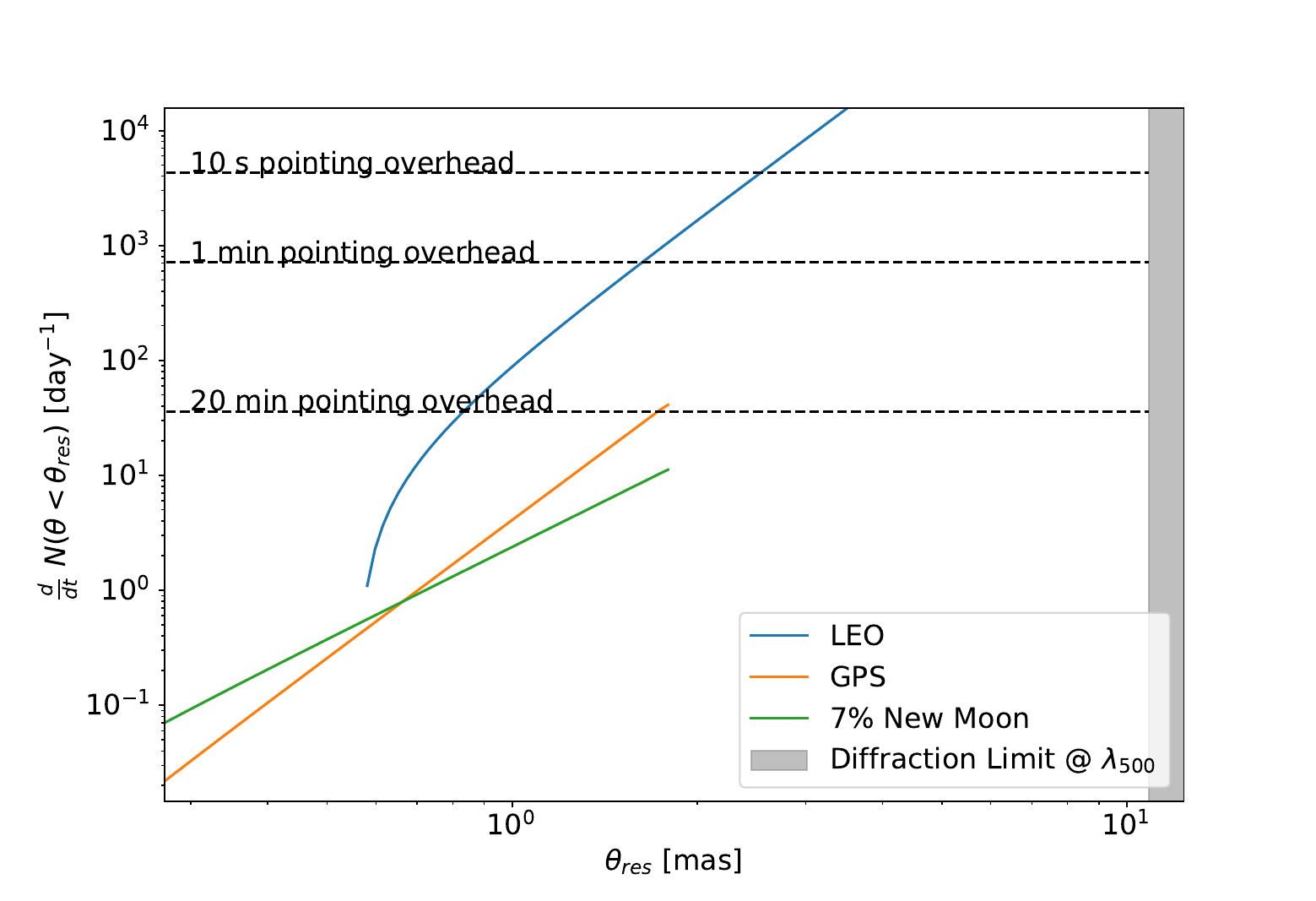}    
    \caption{The cumulative number of events observable every day as a function of angular resolution $\theta_{\rm res}$, assuming the experiment is done on a telescope with an effective aperture of $8$ metre. We also show the maximum amount of observation that can be done, to account for the finite switching time between pointings on the telescope which introduces overhead. Note the curve for LEO is truncated on the left, as our simulation stops when the apparent magnitude of the star reaches $0$ mag and breaks the linear response of the photon sensor. The remaining curves are also truncated on the right, for the apparent magnitude of stars is dimmer than $12$ mag, for which the dark noise of the photon sensor start to have a significant effect.
    }
    \label{fig: resolution-efficiency}
\end{figure}

While using the Moon is favourable for the resolution per occultation event, a complementary analysis of surveying efficiency is equally important.
A full analysis of the occultation rate can be performed by analysing real-time orbit tracking data for the satellites.
For a coarse comparison between masks, we describe the quick estimation procedure below.
Such an approximate analysis will provide baselines for assessing how different masks perform, without being overfitted to the current state of the masks (particularly the artificial satellites), which might change following scheduled future upgrades.

To begin, we estimate the number of stars per apparent magnitude bin.
As the occultation method only applies to a relatively narrow range of bright stars with a cut-off at $m_G \lesssim 12$ mag, we can use an empirical power law to estimate the number of stars.
We calibrate this relation using the Gaia catalogue and the results in \cite{gaia-dr2}:
\footnote{We note that the Gaia catalogue is less complete at the brighter end of the luminosity function, as bright stars can saturate the sensor. In the context of this work, stellar diameters are of particular interest, so the usable bright stars must also have reliable parallax estimates. We thus decide to use the luminosity function from the Gaia catalogue regardless.}
\begin{equation}
    \frac{\partial}{\partial \Omega} N(m<m_*) \approx (3.139 \times 10^{-4} )(10^{0.414m_*}-1) \, \textrm{ deg$^{-2}$}
\end{equation}
At the characteristic values of $m<6$ and $m<10$, the cumulative number densities are $0.10$ deg$^{-2}$ and $4.3$ deg$^{-2}$, respectively.

Based on this, we proceed to estimate the solid angles that various masks sweep across the night sky each night.
Assume that the trajectory of the masks can be reasonably approximated by circular orbits around the Earth's centre.
Over a night of $\sim 12$ hours, they can complete several orbits around the Earth, depending on their angular speed.
Upon projection onto the celestial sphere, such trajectories can be modelled as elongated ribbons.
If we further assume that the ribbons do not overlap, the total solid angle swept is simply:
\begin{equation}
    \Omega(T) = \theta_{\rm geo} \cdot \omega \cdot T,
\end{equation}
where $\theta_{\rm geo}$ is the geometric angular size of the masks, which is much larger than the Fresnel scale $\theta_F$, and $T$ is the duration over which the satellites are tracked.
With these components, we can determine the expected number of daily occultation events as a function of the stars' apparent magnitude.
Mathematically, this is expressed as:
\begin{equation}
    \frac{\partial^2}{\partial m \partial t} N_{\rm event}  = N_{\rm mask} \frac{d\Omega}{dt} \frac{\partial }{\partial m} \left( \frac{\partial N}{\partial\Omega} \right).
\end{equation}
We calculate this occultation rate for different mask choices, with their numerical parameters specified in Table~\ref{tab: occultation-mask-param}, and present the results in Fig.~\ref{fig: event-rate}.
While the Moon has a large angular extent, the solid angle $\Omega$ it sweeps through every day is limited by its low angular speed.
Furthermore, as the full Moon is excessively bright and introduces severe Poisson noise, only the new Moon can be used with appropriate filtering; consequently, the effective daily rate is reduced as the new Moon is available for only a few days each month.
In contrast, artificial satellites are much smaller, but the corresponding event rate is significantly bolstered by their high number in orbit and their substantially faster angular speeds.
Consequently, the daily occultation event rate is dominated by LEO satellites, exceeding that of the Moon by $4$ orders of magnitude.
The event rate for GPS/MEO satellites is significantly lower due to their smaller projected angular size and the fact that they are much less abundant than those in LEO.

However, this does not imply that the use of LEO occultations is optimal.
In fact, the discussion regarding Fig.~\ref{fig: res-lim-appertures} demonstrates that LEO resolution is poorer than that of GPS satellites and the Moon, even when applied to the brightest stars.
Consequently, we examine the occultation event rate as a function of the attainable resolution limit.
This requires fixing the telescope aperture size and optical throughput.
In this feasibility study, we assume an $8$-m telescope with a modest $30\%$ throughput. This choice of throughput is dominated by the quantum efficiency of a Silicon-based avalanche diode, which has been demonstrated to be capable of sub-$\mu$s temporal monitoring \cite{lau2023initial}. 
The result is shown in Fig.\ref{fig: resolution-efficiency}.
Indeed, LEO occultations occur frequently and can generate $\gtrsim 10^5$ events for sub-diffraction-limited measurements.
However, considering the realistic overhead for a large telescope to steer between pointings, it is more practical to utilise telescope time for events with the best angular resolution.
If the telescope is designed such that the pointing overhead is kept within a minute, LEO occultations can provide $\sim 700$ measurements per night with $\lesssim 4$ mas resolution.

Alternatively, if we focus on utilising the idle time of a surveying telescope, we aim to take measurements with the best angular resolution.
Under this consideration, GPS-like satellites in MEO and the new Moon are the optimal choices.
GPS occultation events can provide $\sim 10^1$ measurements every night regardless of the lunar phase, with resolutions no worse than $\sim 3$ mas.
We do not show the rate beyond $3$ mas, as it corresponds to an exceptionally low photon flux $n(t) < 1 \, \mu$s$^{-1}$.
In this limit, there will be $\lesssim 10^2$ photons over a Fresnel crossing time $T_F$.
The flux is also comparable to the dark current of modern ultra-fast photon sensors.
In this limit, the achievable resolution will depend strongly on the noise generation processes of the detector, which can introduce temporally correlated noise over the Fresnel crossing time.
Thus, we refrain from drawing conclusions beyond this limit to ensure our results are applicable to generic photon sensors.

Lastly, the event rate scalings of the Moon and artificial satellites with $\theta_{\rm res}$ are also distinctive.
Although the Moon can offer $\sim 10^{-1}$ mas resolution every few weeks ($10^{-1}$ day$^{-1}$) and is more efficient than GPS satellites, the corresponding event rate drops much faster as Poissonian noise from the Moon becomes comparable to the light from dimmer stars.
In this regime, GPS occultations outperform the Moon; yet LEO occultations also provide coverage in this regime with an event rate an order of magnitude higher.

\section{Conclusion}
\label{sect: conclusion}

We explore the effect of wave-optical occultation of background stars by foreground objects as a mechanism for achieving super-resolution.
Specifically, we examine the use of Low-Earth Orbit (LEO) satellites, GPS-like satellites in the Medium-Earth Orbit (MEO), and the Moon as foreground masks.
We show that foreground objects with approximately Keplerian orbits around the Earth, despite their higher angular speeds, can still produce detectable flux variations because of their larger Fresnel sizes, $\theta_F$.
The typical timescale for light-curve variation is $T_F \sim 10^2 \, \mu$s; determining the stellar angular size requires accurately measuring the light-curve damping, ideally with a temporal resolution that is 2 orders faster than $T_F$, which is $\sim \mathcal{O}(1) \, \mu$s.
We propose a Bayesian framework to model Poissonian photon-counting noise and describe how parameter estimation can properly account for non-Gaussianity in such a photon-starved regime.
Using this tool, we simulate occultation events with various masks and demonstrate how stellar angular sizes can be recovered from the light curves.
The inference pipeline is verified to be capable of recovering injected values, and we estimate the achievable resolution limits in various settings.
In realistic settings, all the masks studied provide angular diameter measurements superior to the optical diffraction limit.
Resolution is primarily limited by the Fresnel size of the foreground masks and photon shot noise.
Regarding attainable angular resolution, lunar occultation is most effective for the brightest stars, achieving $\theta_{\rm res} \sim 10^{-1}$ mas.
However, its resolution degrades rapidly for dimmer stars, eventually being outperformed by GPS satellites when the apparent magnitude is $\gtrsim 8$ mag, owing to excessive photon shot noise from the highly luminous Moon.
LEO satellites can also deliver sub-diffraction-limited measurements, although they offer inferior per-event resolution.
Nonetheless, LEO satellites can provide efficient measurements with an intermediate resolution of $\theta_{\rm res} \sim \mathcal{O}(1)$ mas, offering $N \sim 10^3$ measurements per night, the survey efficiency is mostly limited by telescope pointing overheads.
The Moon and GPS satellites can provide significantly better angular resolution, $\theta_{\rm res} \sim 10^{-1}$ mas, but only at a rate of $N \sim 10^{-1}$ per night, or equivalently once a week.

Based on these findings, we conclude that the Moon and GPS satellites are ideal for studying individual stars with the highest resolution.
However, due to the (pseudo-)random nature of occultation coincidences, it is not possible to select specific scientifically valuable stars for measurement.
Alternatively, LEO satellite occultations provide a compelling case for population-level stellar analysis due to their much higher measurement efficiency.
While the attainable angular resolution is insufficient for resolving most stars, these measurements can still constrain, for example, the ratio of binary to single-star systems.
With such sub-diffractive resolution, the angular diameters of some of the brightest supergiant stars can also be measured by combining this technique with astrometry from Gaia.

As a closing remark, the occultation-based method is primarily limited by photon shot noise.
Unlike diffraction-limited imaging, light curves from multiple occultation events can be stacked, assuming the background stars involved in each event share similar diameters.
Such stacking can improve the signal-to-noise ratio, thereby offering tighter constraints on stellar diameters.
The prospects for stacking analysis and associated scientific use cases are currently undergoing active investigation (Fung et al. \textit{in prep.}).
With recent reports of successful ultra-fast telescope implementations and their associated on-sky testing, we foresee that such measurements can be performed in the very near future.

\appendix
\subsection*{Disclosures}
The authors declare that there are no financial interests, commercial affiliations, or other potential conflicts of interest that could have influenced the objectivity of this research or the writing of this paper.

\subsection*{Code, Data, and Materials Availability} 

The source code for performing the calculation will be shared on reasonable request to the corresponding author.

\subsection* {Acknowledgments}
The authors acknowledge the very valuable feedback from Greg Aldering on the historical context of the method, which is helpful in refining the review section in this revised manuscript.

Leo Fung and Richard Massey acknowledge funding from the UK Science and Technology Facilities Council via grant ST/X001075/1. Fung is a Beloe Scholar who is also partially supported by the \textit{Worshipful Company of Scientific Instrument Makers}.

Albert Wai Kit Lau is a Dunlap postdoctoral fellow; The Dunlap Institute is funded through an endowment established by the David Dunlap family and the University of Toronto.

Leo Fung and Albert Lau would like to dedicate this work to the late George Smoot for his leadership in the SPINA experiment, and more importantly his heartful mentorship and generous support over their early academic careers.

\bibliographystyle{ieeetr}
\bibliography{reference,ref-occultations,ref-stat,ref-spina,ref-superes}

@ARTICLE{gaia-dr2,
       author = {{Gaia Collaboration} and {Brown}, A.~G.~A. and {Vallenari}, A. and {Prusti}, T. and {de Bruijne}, J.~H.~J. and {Babusiaux}, C. and {Bailer-Jones}, C.~A.~L. and {Biermann}, M. and {Evans}, D.~W. and {Eyer}, L. and {Jansen}, F. and {Jordi}, C. and {Klioner}, S.~A. and {Lammers}, U. and {Lindegren}, L. and {Luri}, X. and {Mignard}, F. and {Panem}, C. and {Pourbaix}, D. and {Randich}, S. and {Sartoretti}, P. and {Siddiqui}, H.~I. and {Soubiran}, C. and {van Leeuwen}, F. and {Walton}, N.~A. and {Arenou}, F. and {Bastian}, U. and {Cropper}, M. and {Drimmel}, R. and {Katz}, D. and {Lattanzi}, M.~G. and {Bakker}, J. and {Cacciari}, C. and {Casta{\~n}eda}, J. and {Chaoul}, L. and {Cheek}, N. and {De Angeli}, F. and {Fabricius}, C. and {Guerra}, R. and {Holl}, B. and {Masana}, E. and {Messineo}, R. and {Mowlavi}, N. and {Nienartowicz}, K. and {Panuzzo}, P. and {Portell}, J. and {Riello}, M. and {Seabroke}, G.~M. and {Tanga}, P. and {Th{\'e}venin}, F. and {Gracia-Abril}, G. and {Comoretto}, G. and {Garcia-Reinaldos}, M. and {Teyssier}, D. and {Altmann}, M. and {Andrae}, R. and {Audard}, M. and {Bellas-Velidis}, I. and {Benson}, K. and {Berthier}, J. and {Blomme}, R. and {Burgess}, P. and {Busso}, G. and {Carry}, B. and {Cellino}, A. and {Clementini}, G. and {Clotet}, M. and {Creevey}, O. and {Davidson}, M. and {De Ridder}, J. and {Delchambre}, L. and {Dell'Oro}, A. and {Ducourant}, C. and {Fern{\'a}ndez-Hern{\'a}ndez}, J. and {Fouesneau}, M. and {Fr{\'e}mat}, Y. and {Galluccio}, L. and {Garc{\'\i}a-Torres}, M. and {Gonz{\'a}lez-N{\'u}{\~n}ez}, J. and {Gonz{\'a}lez-Vidal}, J.~J. and {Gosset}, E. and {Guy}, L.~P. and {Halbwachs}, J.-L. and {Hambly}, N.~C. and {Harrison}, D.~L. and {Hern{\'a}ndez}, J. and {Hestroffer}, D. and {Hodgkin}, S.~T. and {Hutton}, A. and {Jasniewicz}, G. and {Jean-Antoine-Piccolo}, A. and {Jordan}, S. and {Korn}, A.~J. and {Krone-Martins}, A. and {Lanzafame}, A.~C. and {Lebzelter}, T. and {L{\"o}ffler}, W. and {Manteiga}, M. and {Marrese}, P.~M. and {Mart{\'\i}n-Fleitas}, J.~M. and {Moitinho}, A. and {Mora}, A. and {Muinonen}, K. and {Osinde}, J. and {Pancino}, E. and {Pauwels}, T. and {Petit}, J.-M. and {Recio-Blanco}, A. and {Richards}, P.~J. and {Rimoldini}, L. and {Robin}, A.~C. and {Sarro}, L.~M. and {Siopis}, C. and {Smith}, M. and {Sozzetti}, A. and {S{\"u}veges}, M. and {Torra}, J. and {van Reeven}, W. and {Abbas}, U. and {Abreu Aramburu}, A. and {Accart}, S. and {Aerts}, C. and {Altavilla}, G. and {{\'A}lvarez}, M.~A. and {Alvarez}, R. and {Alves}, J. and {Anderson}, R.~I. and {Andrei}, A.~H. and {Anglada Varela}, E. and {Antiche}, E. and {Antoja}, T. and {Arcay}, B. and {Astraatmadja}, T.~L. and {Bach}, N. and {Baker}, S.~G. and {Balaguer-N{\'u}{\~n}ez}, L. and {Balm}, P. and {Barache}, C. and {Barata}, C. and {Barbato}, D. and {Barblan}, F. and {Barklem}, P.~S. and {Barrado}, D. and {Barros}, M. and {Barstow}, M.~A. and {Bartholom{\'e} Mu{\~n}oz}, S. and {Bassilana}, J.-L. and {Becciani}, U. and {Bellazzini}, M. and {Berihuete}, A. and {Bertone}, S. and {Bianchi}, L. and {Bienaym{\'e}}, O. and {Blanco-Cuaresma}, S. and {Boch}, T. and {Boeche}, C. and {Bombrun}, A. and {Borrachero}, R. and {Bossini}, D. and {Bouquillon}, S. and {Bourda}, G. and {Bragaglia}, A. and {Bramante}, L. and {Breddels}, M.~A. and {Bressan}, A. and {Brouillet}, N. and {Br{\"u}semeister}, T. and {Brugaletta}, E. and {Bucciarelli}, B. and {Burlacu}, A. and {Busonero}, D. and {Butkevich}, A.~G. and {Buzzi}, R. and {Caffau}, E. and {Cancelliere}, R. and {Cannizzaro}, G. and {Cantat-Gaudin}, T. and {Carballo}, R. and {Carlucci}, T. and {Carrasco}, J.~M. and {Casamiquela}, L. and {Castellani}, M. and {Castro-Ginard}, A. and {Charlot}, P. and {Chemin}, L. and {Chiavassa}, A. and {Cocozza}, G. and {Costigan}, G. and {Cowell}, S. and {Crifo}, F. and {Crosta}, M. and {Crowley}, C. and {Cuypers}, J. and {Dafonte}, C. and {Damerdji}, Y. and {Dapergolas}, A. and {David}, P. and {David}, M. and {de Laverny}, P. and {De Luise}, F.},
        title = "{Gaia Data Release 2. Summary of the contents and survey properties}",
      journal = {\aap},
         year = 2018,
        month = aug,
       volume = {616},
          eid = {A1},
        pages = {A1},
          doi = {10.1051/0004-6361/201833051},
archivePrefix = {arXiv},
       eprint = {1804.09365},
 primaryClass = {astro-ph.GA},
       adsurl = {https://ui.adsabs.harvard.edu/abs/2018A&A...616A...1G}
}

@ARTICLE{lunar-occultation-thai-telescope,
       author = {{Richichi}, A. and {Tasuya}, O. and {Irawati}, P. and {Soonthornthum}, B. and {Dhillon}, V.~S. and {Marsh}, T.~R.},
        title = "{Lunar Occultations of Eighteen Stellar Sources from the 2.4-m Thai National Telescope}",
      journal = {arXiv e-prints},
         year = 2015,
        month = dec,
          eid = {arXiv:1512.06506},
        pages = {arXiv:1512.06506},
          doi = {10.48550/arXiv.1512.06506},
archivePrefix = {arXiv},
       eprint = {1512.06506},
 primaryClass = {astro-ph.SR},
       adsurl = {https://ui.adsabs.harvard.edu/abs/2015arXiv151206506R}
}

@ARTICLE{eso-lunar-occultation-2010,
       author = {{Richichi}, A. and {Fors}, O. and {Chen}, W.-P. and {Mason}, E.},
        title = "{New high-sensitivity, milliarcsecond resolution results from routine observations of lunar occultations at the ESO VLT}",
      journal = {\aap},
         year = 2010,
        month = nov,
       volume = {522},
          eid = {A65},
        pages = {A65},
          doi = {10.1051/0004-6361/201015325},
archivePrefix = {arXiv},
       eprint = {1007.2611},
 primaryClass = {astro-ph.SR},
       adsurl = {https://ui.adsabs.harvard.edu/abs/2010A&A...522A..65R}
}

@ARTICLE{eso-lunar-occultation-catalog2011,
       author = {{Richichi}, A. and {Chen}, W.~P. and {Fors}, O. and {Wang}, P.~F.},
        title = "{Lunar occultations of 184 stellar sources in two crowded regions toward the Galactic bulge}",
      journal = {\aap},
         year = 2011,
        month = aug,
       volume = {532},
          eid = {A101},
        pages = {A101},
          doi = {10.1051/0004-6361/201117282},
archivePrefix = {arXiv},
       eprint = {1105.4816},
 primaryClass = {astro-ph.SR},
       adsurl = {https://ui.adsabs.harvard.edu/abs/2011A&A...532A.101R}
}

@ARTICLE{eso-lunar-occultation-firstRun2008,
       author = {{Richichi}, A. and {Fors}, O. and {Mason}, E. and {Stegmeier}, J. and {Chandrasekhar}, T.},
        title = "{Milliarcsecond angular resolution of reddened stellar sources in the vicinity of the Galactic center}",
      journal = {\aap},
         year = 2008,
        month = oct,
       volume = {489},
       number = {3},
        pages = {1399-1408},
          doi = {10.1051/0004-6361:200810304},
archivePrefix = {arXiv},
       eprint = {0807.2646},
 primaryClass = {astro-ph},
       adsurl = {https://ui.adsabs.harvard.edu/abs/2008A&A...489.1399R}
}

@ARTICLE{lunar-occultation-obs-1987,
       author = {{Stecklum}, Bringfried},
        title = "{Photoelectric Observations of Lunar Occultations}",
      journal = {\aj},
         year = 1987,
        month = jul,
       volume = {94},
        pages = {201},
          doi = {10.1086/114464},
       adsurl = {https://ui.adsabs.harvard.edu/abs/1987AJ.....94..201S}
}

@ARTICLE{asteroid-occultation-2024-sCMOS,
       author = {{Hitchcock}, James and {Gomer}, Richard H.},
        title = "{High-speed imaging system to detect stellar occultations by Kuiper belt and Oort cloud objects}",
      journal = {Journal of Astronomical Telescopes, Instruments, and Systems},
         year = 2024,
        month = jan,
       volume = {10},
          eid = {016003},
        pages = {016003},
          doi = {10.1117/1.JATIS.10.1.016003},
       adsurl = {https://ui.adsabs.harvard.edu/abs/2024JATIS..10a6003H}
}

@ARTICLE{asteroid-occultation-obs-2009-ccd,
       author = {{Bianco}, F.~B. and {Protopapas}, P. and {McLeod}, B.~A. and {Alcock}, C.~R. and {Holman}, M.~J. and {Lehner}, M.~J.},
        title = "{A Search for Occultations of Bright Stars by Small Kuiper Belt Objects Using Megacam on the MMT}",
      journal = {\aj},
         year = 2009,
        month = aug,
       volume = {138},
       number = {2},
        pages = {568-578},
          doi = {10.1088/0004-6256/138/2/568},
archivePrefix = {arXiv},
       eprint = {0903.3036},
 primaryClass = {astro-ph.EP},
       adsurl = {https://ui.adsabs.harvard.edu/abs/2009AJ....138..568B}
}

@ARTICLE{lunar-occultation-obs-1992,
       author = {{Richichi}, A. and {Lisi}, F. and {di Giacomo}, A.},
        title = "{Lunar occultations of southern near-infrared stellar sources.}",
      journal = {\aap},
         year = 1992,
        month = feb,
       volume = {254},
        pages = {149-166},
       adsurl = {https://ui.adsabs.harvard.edu/abs/1992A&A...254..149R}
}

@ARTICLE{starlink-astro-overview,
       author = {{McDowell}, Jonathan C.},
        title = "{The Low Earth Orbit Satellite Population and Impacts of the SpaceX Starlink Constellation}",
      journal = {\apjl},
         year = 2020,
        month = apr,
       volume = {892},
       number = {2},
          eid = {L36},
        pages = {L36},
          doi = {10.3847/2041-8213/ab8016},
archivePrefix = {arXiv},
       eprint = {2003.07446},
 primaryClass = {astro-ph.IM},
       adsurl = {https://ui.adsabs.harvard.edu/abs/2020ApJ...892L..36M}
}

@ARTICLE{occultation-straight-edge-derive,
       author = {{Jennings}, Johnny K. and {McGruder}, III, Charles H.},
        title = "{Comparison of the Disk Diffraction Pattern with the Straight-Edge Diffraction Pattern in Occultations}",
      journal = {\aj},
         year = 1999,
        month = dec,
       volume = {118},
       number = {6},
        pages = {3061-3067},
          doi = {10.1086/301122},
       adsurl = {https://ui.adsabs.harvard.edu/abs/1999AJ....118.3061J}
}

@ARTICLE{whilfford1947-first-lunar-occultation,
       author = {{Whifford}, A.~E.},
        title = "{Angular diameters of stars from occultations by the moon.}",
      journal = {\aj},
         year = 1947,
        month = jan,
       volume = {52},
        pages = {131},
          doi = {10.1086/106023},
       adsurl = {https://ui.adsabs.harvard.edu/abs/1947AJ.....52R.131W}
}

@article{evans-1955-lunar-occultation-analog,
    author = {Evans, David S.},
    title = {An Occultation of Antares Observed at the Union Observatory},
    journal = {Monthly Notices of the Royal Astronomical Society},
    volume = {115},
    number = {5},
    pages = {468-472},
    year = {1955},
    month = {10},
    issn = {0035-8711},
    doi = {10.1093/mnras/115.5.468},
    url = {https://doi.org/10.1093/mnras/115.5.468},
    eprint = {https://academic.oup.com/mnras/article-pdf/115/5/468/9404576/mnras115-0468.pdf},
}

@ARTICLE{lunar-occultation-early-review-1983,
       author = {{White}, N.~M.},
        title = "{Review of the occultation technique for the study of binaries}",
      journal = {Lowell Observatory Bulletin},
         year = 1983,
        month = jan,
       volume = {167},
        pages = {60},
       adsurl = {https://ui.adsabs.harvard.edu/abs/1983LowOB...9...60W}
}

@ARTICLE{spina-noise-model,
       author = {{Lau}, Albert W.~K. and {Fung}, Leo W.~H. and {Smoot}, George F.},
        title = "{False alarm rate-based statistical detection limit for astronomical photon detectors}",
      journal = {Journal of Astronomical Telescopes, Instruments, and Systems},
         year = 2025,
        month = apr,
       volume = {11},
          eid = {028007},
        pages = {028007},
          doi = {10.1117/1.JATIS.11.2.028007},
archivePrefix = {arXiv},
       eprint = {2409.15536},
 primaryClass = {astro-ph.IM},
       adsurl = {https://ui.adsabs.harvard.edu/abs/2025JATIS..11b8007L}
}

@INPROCEEDINGS{iqueye-intensity-interferometry,
       author = {{Zampieri}, Luca and {Naletto}, Giampiero and {Barbieri}, Cesare and {Barbieri}, Mauro and {Verroi}, Enrico and {Umbriaco}, Gabriele and {Favazza}, Paolo and {Lessio}, Luigi and {Farisato}, Giancarlo},
        title = "{Intensity interferometry with Aqueye+ and Iqueye in Asiago}",
    booktitle = {Optical and Infrared Interferometry and Imaging V},
         year = 2016,
       editor = {{Malbet}, Fabien and {Creech-Eakman}, Michelle J. and {Tuthill}, Peter G.},
       series = {Society of Photo-Optical Instrumentation Engineers (SPIE) Conference Series},
       volume = {9907},
        month = aug,
          eid = {99070N},
        pages = {99070N},
          doi = {10.1117/12.2233688},
archivePrefix = {arXiv},
       eprint = {1609.01134},
 primaryClass = {astro-ph.IM},
       adsurl = {https://ui.adsabs.harvard.edu/abs/2016SPIE.9907E..0NZ}
}

@article{lau2020sky,
  title={On-sky silicon photomultiplier detector performance measurements for millisecond to sub-microsecond optical source variability studies},
  author={Lau, Albert WK and Shafiee, Mehdi and Smoot, George F and Grossan, Bruce and Li, Siyang and Maksut, Zhanat},
  journal={Journal of Astronomical Telescopes, Instruments, and Systems},
  volume={6},
  number={4},
  pages={046002--046002},
  year={2020},
  publisher={Society of Photo-Optical Instrumentation Engineers}
}

@inproceedings{lau2022development,
  title={Development of position-sensitive photon-counting imager for Ultra-Fast Astronomy},
  author={Lau, Albert Wai Kit and Chan, Yan Yan and Shafiee, Mehdi and Smoot, George F and Grossan, Bruce},
  booktitle={X-Ray, Optical, and Infrared Detectors for Astronomy X},
  volume={12191},
  pages={312--329},
  year={2022},
  organization={SPIE}
}

@article{lau2023initial,
  title={Initial On-Sky Performance Testing of the Single-Photon Imager for Nanosecond Astrophysics (SPINA) System},
  author={Lau, Albert Wai Kit and Shaimoldin, Nurzhan and Maksut, Zhanat and Chan, Yan Yan and Shafiee, Mehdi and Grossan, Bruce and Smoot, George F},
  journal={IEEE Transactions on Instrumentation and Measurement},
  year={2023},
  publisher={IEEE}
}

@article{sr-sim-review,
author = {Heintzmann, Rainer and Huser, Thomas},
title = {Super-Resolution Structured Illumination Microscopy},
journal = {Chemical Reviews},
volume = {117},
number = {23},
pages = {13890-13908},
year = {2017},
doi = {10.1021/acs.chemrev.7b00218},
    note ={PMID: 29125755},

eprint = { 
        https://doi.org/10.1021/acs.chemrev.7b00218
}
}

@ARTICLE{sr-sim-review-2021,
       author = {{Samanta}, Krishnendu and {Joseph}, Joby},
        title = "{An overview of structured illumination microscopy: recent advances and perspectives}",
      journal = {Journal of Optics},
         year = 2021,
        month = dec,
       volume = {23},
       number = {12},
          eid = {123002},
        pages = {123002},
          doi = {10.1088/2040-8986/ac3675},
       adsurl = {https://ui.adsabs.harvard.edu/abs/2021JOpt...23l3002S}
}

@ARTICLE{structured-illumination-origin-idea-1993,
       author = {{Bailey}, Brent and {Farkas}, Daniel L. and {Taylor}, D. Lansing and {Lanni}, Frederick},
        title = "{Enhancement of axial resolution in fluorescence microscopy by standing-wave excitation}",
      journal = {Nature},
         year = 1993,
        month = nov,
       volume = {366},
       number = {6450},
        pages = {44-48},
          doi = {10.1038/366044a0},
       adsurl = {https://ui.adsabs.harvard.edu/abs/1993Natur.366...44B}
}

@article{structured-illumination-first-implement-2000,
author = {Gustafsson, M. G. L.},
title = {Surpassing the lateral resolution limit by a factor of two using structured illumination microscopy},
journal = {Journal of Microscopy},
volume = {198},
number = {2},
pages = {82-87},
doi = {https://doi.org/10.1046/j.1365-2818.2000.00710.x},

eprint = {https://onlinelibrary.wiley.com/doi/pdf/10.1046/j.1365-2818.2000.00710.x},
year = {2000}
}

@article{STED-original-hell94,
author = {Stefan W. Hell and Jan Wichmann},
journal = {Opt. Lett.},
number = {11},
pages = {780--782},
publisher = {Optica Publishing Group},
title = {Breaking the diffraction resolution limit by stimulated emission: stimulated-emission-depletion fluorescence microscopy},
volume = {19},
month = {Jun},
year = {1994},

doi = {10.1364/OL.19.000780},
}

@article{STED-review-2017,
author = {Blom, Hans and Widengren, Jerker},
title = {Stimulated Emission Depletion Microscopy},
journal = {Chemical Reviews},
volume = {117},
number = {11},
pages = {7377-7427},
year = {2017},
doi = {10.1021/acs.chemrev.6b00653},
    note ={PMID: 28262022},


eprint = { 
    
        https://doi.org/10.1021/acs.chemrev.6b00653
    
    

}

}

@ARTICLE{darkmoon-flux-obs,
       author = {{Thejll}, P. and {Flynn}, C. and {Gleisner}, H. and {Andersen}, T. and {Ulla}, A. and {O-Petersen}, M. and {Darudi}, A. and {Schwarz}, H.},
        title = "{The colour of the dark side of the Moon}",
      journal = {\aap},
         year = 2014,
        month = mar,
       volume = {563},
          eid = {A38},
        pages = {A38},
          doi = {10.1051/0004-6361/201322776},
archivePrefix = {arXiv},
       eprint = {1401.1994},
 primaryClass = {astro-ph.EP},
       adsurl = {https://ui.adsabs.harvard.edu/abs/2014A&A...563A..38T}
}

@article{meinel-telescope-cost-scaling,
author = {Aden B. Meinel},
title = {{Cost-Scaling Laws Applicable To Very Large Optical Telescopes}},
volume = {18},
journal = {Optical Engineering},
number = {6},
publisher = {SPIE},
pages = {186645},
year = {1979},
doi = {10.1117/12.7972448},
URL = {https://doi.org/10.1117/12.7972448}
}

@article{adaptive-optics-ELT-algo,
    author = {Basden, A G and Jenkins, D and Morris, T J and Osborn, J and Townson, M J},
    title = {Efficient implementation of pseudo open-loop control for adaptive optics on Extremely Large Telescopes},
    journal = {Monthly Notices of the Royal Astronomical Society},
    volume = {486},
    number = {2},
    pages = {1774-1780},
    year = {2019},
    month = {03},
   
    issn = {0035-8711},
    doi = {10.1093/mnras/stz918},
    url = {https://doi.org/10.1093/mnras/stz918},
    eprint = {https://academic.oup.com/mnras/article-pdf/486/2/1774/28484708/stz918.pdf},
}

@ARTICLE{veritas-intensity-inteferometry,
       author = {{Abeysekara}, A.~U. and {Benbow}, W. and {Brill}, A. and {Buckley}, J.~H. and {Christiansen}, J.~L. and {Chromey}, A.~J. and {Daniel}, M.~K. and {Davis}, J. and {Falcone}, A. and {Feng}, Q. and {Finley}, J.~P. and {Fortson}, L. and {Furniss}, A. and {Gent}, A. and {Giuri}, C. and {Gueta}, O. and {Hanna}, D. and {Hassan}, T. and {Hervet}, O. and {Holder}, J. and {Hughes}, G. and {Humensky}, T.~B. and {Kaaret}, P. and {Kertzman}, M. and {Kieda}, D. and {Krennrich}, F. and {Kumar}, S. and {LeBohec}, T. and {Lin}, T.~T.~Y. and {Lundy}, M. and {Maier}, G. and {Matthews}, N. and {Moriarty}, P. and {Mukherjee}, R. and {Nievas-Rosillo}, M. and {O'Brien}, S. and {Ong}, R.~A. and {Otte}, A.~N. and {Pfrang}, K. and {Pohl}, M. and {Prado}, R.~R. and {Pueschel}, E. and {Quinn}, J. and {Ragan}, K. and {Reynolds}, P.~T. and {Ribeiro}, D. and {Richards}, G.~T. and {Roache}, E. and {Ryan}, J.~L. and {Santander}, M. and {Sembroski}, G.~H. and {Wakely}, S.~P. and {Weinstein}, A. and {Wilcox}, P. and {Williams}, D.~A. and {Williamson}, T.~J.},
        title = "{Demonstration of stellar intensity interferometry with the four VERITAS telescopes}",
      journal = {Nature Astronomy},
         year = 2020,
        month = jan,
       volume = {4},
        pages = {1164-1169},
          doi = {10.1038/s41550-020-1143-y},
archivePrefix = {arXiv},
       eprint = {2007.10295},
 primaryClass = {astro-ph.IM},
       adsurl = {https://ui.adsabs.harvard.edu/abs/2020NatAs...4.1164A}
}

@ARTICLE{veritas-asteroid-occultation,
       author = {{Benbow}, W. and {Bird}, R. and {Brill}, A. and {Brose}, R. and {Chromey}, A.~J. and {Daniel}, M.~K. and {Feng}, Q. and {Finley}, J.~P. and {Fortson}, L. and {Furniss}, A. and {Gillanders}, G.~H. and {Giuri}, C. and {Gueta}, O. and {Hanna}, D. and {Halpern}, J.~P. and {Hassan}, T. and {Holder}, J. and {Hughes}, G. and {Humensky}, T.~B. and {Joyce}, A.~M. and {Kaaret}, P. and {Kar}, P. and {Kelley-Hoskins}, N. and {Kertzman}, M. and {Kieda}, D. and {Krause}, M. and {Lang}, M.~J. and {Lin}, T.~T.~Y. and {Maier}, G. and {Matthews}, N. and {Moriarty}, P. and {Mukherjee}, R. and {Nieto}, D. and {Nievas-Rosillo}, M. and {O'Brien}, S. and {Ong}, R.~A. and {Park}, N. and {Petrashyk}, A. and {Pohl}, M. and {Pueschel}, E. and {Quinn}, J. and {Ragan}, K. and {Reynolds}, P.~T. and {Richards}, G.~T. and {Roache}, E. and {Rulten}, C. and {Sadeh}, I. and {Santander}, M. and {Sembroski}, G.~H. and {Shahinyan}, K. and {Sushch}, I. and {Wakely}, S.~P. and {Wells}, R.~M. and {Wilcox}, P. and {Wilhelm}, A. and {Williams}, D.~A. and {Williamson}, T.~J.},
        title = "{Direct measurement of stellar angular diameters by the VERITAS Cherenkov telescopes}",
      journal = {Nature Astronomy},
         year = 2019,
        month = apr,
       volume = {3},
        pages = {511-516},
          doi = {10.1038/s41550-019-0741-z},
archivePrefix = {arXiv},
       eprint = {1904.06324},
 primaryClass = {astro-ph.SR},
       adsurl = {https://ui.adsabs.harvard.edu/abs/2019NatAs...3..511B}
}

@ARTICLE{hbt-effect-original1956,
       author = {{Brown}, R. Hanbury and {Twiss}, R.~Q.},
        title = "{Correlation between Photons in two Coherent Beams of Light}",
      journal = {\nat},
         year = 1956,
        month = jan,
       volume = {177},
       number = {4497},
        pages = {27-29},
          doi = {10.1038/177027a0},
       adsurl = {https://ui.adsabs.harvard.edu/abs/1956Natur.177...27B}
}

@ARTICLE{stellar-interferometer-original1956,
       author = {{Hanbury Brown}, R.},
        title = "{A Test of a New Type of Stellar Interferometer on Sirius}",
      journal = {\nat},
         year = 1956,
        month = nov,
       volume = {178},
       number = {4541},
        pages = {1046-1048},
          doi = {10.1038/1781046a0},
       adsurl = {https://ui.adsabs.harvard.edu/abs/1956Natur.178.1046H}
}

@BOOK{cta-science-paper,
       author = {{Cherenkov Telescope Array Consortium} and {Acharya}, B.~S. and {Agudo}, I. and {Al Samarai}, I. and {Alfaro}, R. and {Alfaro}, J. and {Alispach}, C. and {Alves Batista}, R. and {Amans}, J.-P. and {Amato}, E. and {Ambrosi}, G. and {Antolini}, E. and {Antonelli}, L.~A. and {Aramo}, C. and {Araya}, M. and {Armstrong}, T. and {Arqueros}, F. and {Arrabito}, L. and {Asano}, K. and {Ashley}, M. and {Backes}, M. and {Balazs}, C. and {Balbo}, M. and {Ballester}, O. and {Ballet}, J. and {Bamba}, A. and {Barkov}, M. and {Barres de Almeida}, U. and {Barrio}, J.~A. and {Bastieri}, D. and {Becherini}, Y. and {Belfiore}, A. and {Benbow}, W. and {Berge}, D. and {Bernardini}, E. and {Bernardini}, M.~G. and {Bernardos}, M. and {Bernl{\"o}hr}, K. and {Bertucci}, B. and {Biasuzzi}, B. and {Bigongiari}, C. and {Biland}, A. and {Bissaldi}, E. and {Biteau}, J. and {Blanch}, O. and {Blazek}, J. and {Boisson}, C. and {Bolmont}, J. and {Bonanno}, G. and {Bonardi}, A. and {Bonavolont{\`a}}, C. and {Bonnoli}, G. and {Bosnjak}, Z. and {B{\"o}ttcher}, M. and {Braiding}, C. and {Bregeon}, J. and {Brill}, A. and {Brown}, A.~M. and {Brun}, P. and {Brunetti}, G. and {Buanes}, T. and {Buckley}, J. and {Bugaev}, V. and {B{\"u}hler}, R. and {Bulgarelli}, A. and {Bulik}, T. and {Burton}, M. and {Burtovoi}, A. and {Busetto}, G. and {Canestrari}, R. and {Capalbi}, M. and {Capitanio}, F. and {Caproni}, A. and {Caraveo}, P. and {C{\'a}rdenas}, V. and {Carlile}, C. and {Carosi}, R. and {Carqu{\'\i}n}, E. and {Carr}, J. and {Casanova}, S. and {Cascone}, E. and {Catalani}, F. and {Catalano}, O. and {Cauz}, D. and {Cerruti}, M. and {Chadwick}, P. and {Chaty}, S. and {Chaves}, R.~C.~G. and {Chen}, A. and {Chen}, X. and {Chernyakova}, M. and {Chikawa}, M. and {Christov}, A. and {Chudoba}, J. and {Cie{\'s}lar}, M. and {Coco}, V. and {Colafrancesco}, S. and {Colin}, P. and {Conforti}, V. and {Connaughton}, V. and {Conrad}, J. and {Contreras}, J.~L. and {Cortina}, J. and {Costa}, A. and {Costantini}, H. and {Cotter}, G. and {Covino}, S. and {Crocker}, R. and {Cuadra}, J. and {Cuevas}, O. and {Cumani}, P. and {D'A{\`\i}}, A. and {D'Ammando}, F. and {D'Avanzo}, P. and {D'Urso}, D. and {Daniel}, M. and {Davids}, I. and {Dawson}, B. and {Dazzi}, F. and {De Angelis}, A. and {de C{\'a}ssia dos Anjos}, R. and {De Cesare}, G. and {De Franco}, A. and {de Gouveia Dal Pino}, E.~M. and {de la Calle}, I. and {de los Reyes Lopez}, R. and {De Lotto}, B. and {De Luca}, A. and {De Lucia}, M. and {de Naurois}, M. and {de O{\~n}a Wilhelmi}, E. and {De Palma}, F. and {De Persio}, F. and {de Souza}, V. and {Deil}, C. and {Del Santo}, M. and {Delgado}, C. and {della Volpe}, D. and {Di Girolamo}, T. and {Di Pierro}, F. and {Di Venere}, L. and {D{\'\i}az}, C. and {Dib}, C. and {Diebold}, S. and {Djannati-Ata{\"\i}}, A. and {Dom{\'\i}nguez}, A. and {Dominis Prester}, D. and {Dorner}, D. and {Doro}, M. and {Drass}, H. and {Dravins}, D. and {Dubus}, G. and {Dwarkadas}, V.~V. and {Ebr}, J. and {Eckner}, C. and {Egberts}, K. and {Einecke}, S. and {Ekoume}, T.~R.~N. and {Els{\"a}sser}, D. and {Ernenwein}, J.-P. and {Espinoza}, C. and {Evoli}, C. and {Fairbairn}, M. and {Falceta-Goncalves}, D. and {Falcone}, A. and {Farnier}, C. and {Fasola}, G. and {Fedorova}, E. and {Fegan}, S. and {Fernandez-Alonso}, M. and {Fern{\'a}ndez-Barral}, A. and {Ferrand}, G. and {Fesquet}, M. and {Filipovic}, M. and {Fioretti}, V. and {Fontaine}, G. and {Fornasa}, M. and {Fortson}, L. and {Freixas Coromina}, L. and {Fruck}, C. and {Fujita}, Y. and {Fukazawa}, Y. and {Funk}, S. and {F{\"u}{\ss}ling}, M. and {Gabici}, S. and {Gadola}, A. and {Gallant}, Y. and {Garcia}, B. and {Garcia L{\'o}pez}, R. and {Garczarczyk}, M. and {Gaskins}, J. and {Gasparetto}, T. and {Gaug}, M. and {Gerard}, L. and {Giavitto}, G. and {Giglietto}, N. and {Giommi}, P. and {Giordano}, F. and {Giro}, E. and {Giroletti}, M.},
        title = "{Science with the Cherenkov Telescope Array}",
         year = 2019,
          doi = {10.1142/10986},
       adsurl = {https://ui.adsabs.harvard.edu/abs/2019scta.book.....C}
}

@article{adaptive-optics-review,
  title={Adaptive optics for high-resolution imaging},
  author={Hampson, Karen M and Turcotte, Rapha{\"e}l and Miller, Donald T and Kurokawa, Kazuhiro and Males, Jared R and Ji, Na and Booth, Martin J},
  journal={Nature Reviews Methods Primers},
  volume={1},
  number={1},
  pages={68},
  year={2021},
  publisher={Nature Publishing Group UK London}
}

@ARTICLE{adaptive-optics-for-astro-review,
       author = {{Davies}, Richard and {Kasper}, Markus},
        title = "{Adaptive Optics for Astronomy}",
      journal = {\araa},
         year = 2012,
        month = sep,
       volume = {50},
        pages = {305-351},
          doi = {10.1146/annurev-astro-081811-125447},
archivePrefix = {arXiv},
       eprint = {1201.5741},
 primaryClass = {astro-ph.IM},
       adsurl = {https://ui.adsabs.harvard.edu/abs/2012ARA&A..50..305D}
}

@ARTICLE{keck-adaptive-optics,
       author = {{Wizinowich}, P. and {Acton}, D.~S. and {Shelton}, C. and {Stomski}, P. and {Gathright}, J. and {Ho}, K. and {Lupton}, W. and {Tsubota}, K. and {Lai}, O. and {Max}, C. and {Brase}, J. and {An}, J. and {Avicola}, K. and {Olivier}, S. and {Gavel}, D. and {Macintosh}, B. and {Ghez}, A. and {Larkin}, J.},
        title = "{First Light Adaptive Optics Images from the Keck II Telescope: A New Era of High Angular Resolution Imagery}",
      journal = {\pasp},
         year = 2000,
        month = mar,
       volume = {112},
       number = {769},
        pages = {315-319},
          doi = {10.1086/316543},
       adsurl = {https://ui.adsabs.harvard.edu/abs/2000PASP..112..315W}
}

@ARTICLE{subaru-adaptive-optics,
       author = {{Jovanovic}, N. and {Martinache}, F. and {Guyon}, O. and {Clergeon}, C. and {Singh}, G. and {Kudo}, T. and {Garrel}, V. and {Newman}, K. and {Doughty}, D. and {Lozi}, J. and {Males}, J. and {Minowa}, Y. and {Hayano}, Y. and {Takato}, N. and {Morino}, J. and {Kuhn}, J. and {Serabyn}, E. and {Norris}, B. and {Tuthill}, P. and {Schworer}, G. and {Stewart}, P. and {Close}, L. and {Huby}, E. and {Perrin}, G. and {Lacour}, S. and {Gauchet}, L. and {Vievard}, S. and {Murakami}, N. and {Oshiyama}, F. and {Baba}, N. and {Matsuo}, T. and {Nishikawa}, J. and {Tamura}, M. and {Lai}, O. and {Marchis}, F. and {Duchene}, G. and {Kotani}, T. and {Woillez}, J.},
        title = "{The Subaru Coronagraphic Extreme Adaptive Optics System: Enabling High-Contrast Imaging on Solar-System Scales}",
      journal = {\pasp},
         year = 2015,
        month = sep,
       volume = {127},
       number = {955},
        pages = {890},
          doi = {10.1086/682989},
archivePrefix = {arXiv},
       eprint = {1507.00017},
 primaryClass = {astro-ph.IM},
       adsurl = {https://ui.adsabs.harvard.edu/abs/2015PASP..127..890J}
}

@ARTICLE{vera-rubin-adaptive-optics,
       author = {{Megias Homar}, Guillem and {Kahn}, Steven M. and {Meyers}, Joshua M. and {Crenshaw}, John Franklin and {Thomas}, Sandrine J.},
        title = "{The Active Optics System on the Vera C. Rubin Observatory: Optimal Control of Degeneracy among the Large Number of Degrees of Freedom}",
      journal = {\apj},
         year = 2024,
        month = oct,
       volume = {974},
       number = {1},
          eid = {108},
        pages = {108},
          doi = {10.3847/1538-4357/ad6cdc},
archivePrefix = {arXiv},
       eprint = {2406.04656},
 primaryClass = {astro-ph.IM},
       adsurl = {https://ui.adsabs.harvard.edu/abs/2024ApJ...974..108M}
}

@INCOLLECTION{exoplanet-transit-review,
       author = {{Winn}, J.~N.},
        title = "{Exoplanet Transits and Occultations}",
    booktitle = {Exoplanets},
         year = 2010,
       editor = {{Seager}, S.},
        pages = {55-77},
          doi = {10.48550/arXiv.1001.2010},
       adsurl = {https://ui.adsabs.harvard.edu/abs/2010exop.book...55W}
}

@article{stellar-evolution-chandrasekhar-review1984,
  title={On stars, their evolution and their stability},
  author={Chandrasekhar, Subrahmanyan},
  journal={Science},
  volume={226},
  number={4674},
  pages={497--505},
  year={1984},
  publisher={American Association for the Advancement of Science}
}

@article{stellar-structure-evolution-review2015,
  title={Structure and evolution of stars},
  author={Schwarzschild, Martin},
  year={2015},
  publisher={Princeton University Press}
}

@ARTICLE{agn-size-problem-2025,
       author = {{Lewin}, Collin and {Kara}, Erin and {Panagiotou}, Christos and {Cackett}, Edward M. and {Gelbord}, Jonathan and {Hern{\'a}ndez Santisteban}, Juan V. and {Horne}, Keith and {Kriss}, Gerard A.},
        title = "{The Accretion Disk Size Problem in AGN Disk Reverberation Mapping Is an Obscuration Effect: A Uniform AGN Sample Study with Swift}",
      journal = {\apj},
         year = 2025,
        month = nov,
       volume = {993},
       number = {2},
          eid = {245},
        pages = {245},
          doi = {10.3847/1538-4357/ae0b6c},
archivePrefix = {arXiv},
       eprint = {2509.25315},
 primaryClass = {astro-ph.HE},
       adsurl = {https://ui.adsabs.harvard.edu/abs/2025ApJ...993..245L}
}

@ARTICLE{agn-size-reverberation-map,
       author = {{Jha}, Vivek Kumar and {Joshi}, Ravi and {Saraswat}, Jayesh and {Chand}, Hum and {Barway}, Sudhanshu and {Mandal}, Amit Kumar},
        title = "{Exploring the AGN Accretion Disks Using Continuum Reverberation Mapping}",
      journal = {Bulletin de la Societe Royale des Sciences de Liege},
         year = 2024,
        month = jun,
       volume = {93},
       number = {2},
        pages = {766-779},
          doi = {10.25518/0037-9565.11871},
archivePrefix = {arXiv},
       eprint = {2307.16568},
 primaryClass = {astro-ph.HE},
       adsurl = {https://ui.adsabs.harvard.edu/abs/2024BSRSL..93..766J}
}

@ARTICLE{agn-size-reveberation-map-measure,
       author = {{Mandal}, Amit Kumar and {Woo}, Jong-Hak and {Wang}, Shu},
        title = "{The Size of the Continuum Emission Region and Its Scaling Relations with Active Galactic Nucleus Luminosity and the Broad-line Region Size}",
      journal = {\apj},
         year = 2025,
        month = may,
       volume = {985},
       number = {1},
          eid = {30},
        pages = {30},
          doi = {10.3847/1538-4357/adc56e},
archivePrefix = {arXiv},
       eprint = {2502.19184},
 primaryClass = {astro-ph.GA},
       adsurl = {https://ui.adsabs.harvard.edu/abs/2025ApJ...985...30M}
}

@ARTICLE{imbh-with-lensed-gw,
       author = {{Lai}, Kwun-Hang and {Hannuksela}, Otto A. and {Herrera-Mart{\'\i}n}, Antonio and {Diego}, Jose M. and {Broadhurst}, Tom and {Li}, Tjonnie G.~F.},
        title = "{Discovering intermediate-mass black hole lenses through gravitational wave lensing}",
      journal = {\prd},
         year = 2018,
        month = oct,
       volume = {98},
       number = {8},
          eid = {083005},
        pages = {083005},
          doi = {10.1103/PhysRevD.98.083005},
archivePrefix = {arXiv},
       eprint = {1801.07840},
 primaryClass = {gr-qc},
       adsurl = {https://ui.adsabs.harvard.edu/abs/2018PhRvD..98h3005L}
}

@ARTICLE{imbh-detection-globular-lensing,
       author = {{Tatekawa}, Takayuki and {Okamura}, Yuuki},
        title = "{Detection of intermediate-mass black holes in globular clusters using gravitational lensing}",
      journal = {Stars and Galaxies},
         year = 2020,
        month = dec,
       volume = {3},
        pages = {3},
          doi = {10.32231/starsandgalaxies.3.0_3},
archivePrefix = {arXiv},
       eprint = {2012.14703},
 primaryClass = {astro-ph.GA},
       adsurl = {https://ui.adsabs.harvard.edu/abs/2020StGal...3....3T}
}

@article{massive-star-evolution,
  title={The evolution of massive stars with mass loss},
  author={Chiosi, Cesare and Maeder, Andre},
  journal={IN: Annual review of astronomy and astrophysics. Volume 24 (A87-26730 10-90). Palo Alto, CA, Annual Reviews, Inc., 1986, p. 329-375. CNR-SNSF-supported research.},
  volume={24},
  pages={329--375},
  year={1986}
}

@ARTICLE{stellar-intensity-interferometry-magic,
       author = {{Abe}, S. and {Abhir}, J. and {Acciari}, V.~A. and {Aguasca-Cabot}, A. and {Agudo}, I. and {Aniello}, T. and {Ansoldi}, S. and {Antonelli}, L.~A. and {Arbet Engels}, A. and {Arcaro}, C. and {Artero}, M. and {Asano}, K. and {Babi{\'c}}, A. and {Baquero}, A. and {de Almeida}, U. Barres and {Barrio}, J.~A. and {Batkovi{\'c}}, I. and {Bautista}, A. and {Baxter}, J. and {Gonz{\'a}lez}, J. Becerra and {Bernardini}, E. and {Bernardos}, M. and {Bernete}, J. and {Berti}, A. and {Besenrieder}, J. and {Bigongiari}, C. and {Biland}, A. and {Blanch}, O. and {Bonnoli}, G. and {Bo{\v{s}}njak}, {\v{Z}}. and {Burelli}, I. and {Busetto}, G. and {Campoy-Ordaz}, A. and {Carosi}, A. and {Carosi}, R. and {Carretero-Castrillo}, M. and {Ceribella}, G. and {Chai}, Y. and {Cifuentes}, A. and {Colombo}, E. and {Contreras}, J.~L. and {Cortina}, J. and {Covino}, S. and {D'Amico}, G. and {D'Elia}, V. and {Da Vela}, P. and {Dazzi}, F. and {De Angelis}, A. and {De Lotto}, B. and {de Menezes}, R. and {Del Popolo}, A. and {Delfino}, M. and {Delgado}, J. and {Delgado Mendez}, C. and {Di Pierro}, F. and {Di Venere}, L. and {Dominis Prester}, D. and {Donini}, A. and {Dorner}, D. and {Doro}, M. and {Elsaesser}, D. and {Emery}, G. and {Escudero}, J. and {Fari{\~n}a}, L. and {Fattorini}, A. and {Foffano}, L. and {Font}, L. and {Fr{\"o}se}, S. and {Fukami}, S. and {Fukazawa}, Y. and {Garc{\'\i}a L{\'o}pez}, R.~J. and {Garczarczyk}, M. and {Gasparyan}, S. and {Gaug}, M. and {Giesbrecht Paiva}, J.~G. and {Giglietto}, N. and {Giordano}, F. and {Gliwny}, P. and {Gradetzke}, T. and {Grau}, R. and {Green}, D. and {Green}, J.~G. and {G{\"u}nther}, P. and {Hadasch}, D. and {Hahn}, A. and {Hassan}, T. and {Heckmann}, L. and {Herrera}, J. and {Hrupec}, D. and {H{\"u}tten}, M. and {Imazawa}, R. and {Ishio}, K. and {Jim{\'e}nez Mart{\'\i}nez}, I. and {Jormanainen}, J. and {Kayanoki}, T. and {Kerszberg}, D. and {Kluge}, G.~W. and {Kobayashi}, Y. and {Kouch}, P.~M. and {Kubo}, H. and {Kushida}, J. and {L{\'a}inez}, M. and {Lamastra}, A. and {Leone}, F. and {Lindfors}, E. and {Linhoff}, L. and {Lombardi}, S. and {Longo}, F. and {L{\'o}pez-Coto}, R. and {L{\'o}pez-Moya}, M. and {L{\'o}pez-Oramas}, A. and {Loporchio}, S. and {Lorini}, A. and {Lyard}, E. and {Machado de Oliveira Fraga}, B. and {Majumdar}, P. and {Makariev}, M. and {Maneva}, G. and {Mang}, N. and {Manganaro}, M. and {Mangano}, S. and {Mannheim}, K. and {Mariotti}, M. and {Mart{\'\i}nez}, M. and {Mart{\'\i}nez-Chicharro}, M. and {Mas-Aguilar}, A. and {Mazin}, D. and {Menchiari}, S. and {Mender}, S. and {Miceli}, D. and {Miener}, T. and {Miranda}, J.~M. and {Mirzoyan}, R. and {Molero Gonz{\'a}lez}, M. and {Molina}, E. and {Mondal}, H.~A. and {Moralejo}, A. and {Morcuende}, D. and {Nakamori}, T. and {Nanci}, C. and {Neustroev}, V. and {Nickel}, L. and {Nievas Rosillo}, M. and {Nigro}, C. and {Nikoli{\'c}}, L. and {Nilsson}, K. and {Nishijima}, K. and {Ekoume}, T. Njoh and {Noda}, K. and {Nozaki}, S. and {Ohtani}, Y. and {Okumura}, A. and {Otero-Santos}, J. and {Paiano}, S. and {Palatiello}, M. and {Paneque}, D. and {Paoletti}, R. and {Paredes}, J.~M. and {Peresano}, M. and {Persic}, M. and {Pihet}, M. and {Pirola}, G. and {Podobnik}, F. and {Prada Moroni}, P.~G. and {Prandini}, E. and {Principe}, G. and {Priyadarshi}, C. and {Rhode}, W. and {Rib{\'o}}, M. and {Rico}, J. and {Righi}, C. and {Sahakyan}, N. and {Saito}, T. and {Satalecka}, K. and {Saturni}, F.~G. and {Schleicher}, B. and {Schmidt}, K. and {Schmuckermaier}, F. and {Schubert}, J.~L. and {Schweizer}, T. and {Sciaccaluga}, A. and {Silvestri}, G. and {Sitarek}, J. and {Sliusar}, V. and {Sobczynska}, D. and {Spolon}, A. and {Stamerra}, A. and {Stri{\v{s}}kovi{\'c}}, J. and {Strom}, D. and {Strzys}, M. and {Suda}, Y. and {Suri{\'c}}, T. and {Suutarinen}, S. and {Tajima}, H. and {Takahashi}, M. and {Takeishi}, R. and {Temnikov}, P. and {Terauchi}, K. and {Terzi{\'c}}, T. and {Teshima}, M.},
        title = "{Performance and first measurements of the MAGIC stellar intensity interferometer}",
      journal = {\mnras},
         year = 2024,
        month = apr,
       volume = {529},
       number = {4},
        pages = {4387-4404},
          doi = {10.1093/mnras/stae697},
archivePrefix = {arXiv},
       eprint = {2402.04755},
 primaryClass = {astro-ph.IM},
       adsurl = {https://ui.adsabs.harvard.edu/abs/2024MNRAS.529.4387A}
}

@ARTICLE{stellar-intensity-interferometry-2012,
       author = {{Nu{\~n}ez}, Paul D. and {Holmes}, Richard and {Kieda}, David and {Lebohec}, Stephan},
        title = "{High angular resolution imaging with stellar intensity interferometry using air Cherenkov telescope arrays}",
      journal = {\mnras},
         year = 2012,
        month = jan,
       volume = {419},
       number = {1},
        pages = {172-183},
          doi = {10.1111/j.1365-2966.2011.19683.x},
archivePrefix = {arXiv},
       eprint = {1108.4682},
 primaryClass = {astro-ph.IM},
       adsurl = {https://ui.adsabs.harvard.edu/abs/2012MNRAS.419..172N}
}

@ARTICLE{starlink_impact_mroz,
       author = {{Mr{\'o}z}, Przemek and {Otarola}, Angel and {Prince}, Thomas A. and {Dekany}, Richard and {Duev}, Dmitry A. and {Graham}, Matthew J. and {Groom}, Steven L. and {Masci}, Frank J. and {Medford}, Michael S.},
        title = "{Impact of the SpaceX Starlink Satellites on the Zwicky Transient Facility Survey Observations}",
      journal = {\apjl},
         year = 2022,
        month = jan,
       volume = {924},
       number = {2},
          eid = {L30},
        pages = {L30},
          doi = {10.3847/2041-8213/ac470a},
archivePrefix = {arXiv},
       eprint = {2201.05343},
 primaryClass = {astro-ph.IM},
       adsurl = {https://ui.adsabs.harvard.edu/abs/2022ApJ...924L..30M}
}

@ARTICLE{starlink_impact_mendoza,
       author = {{Serrano Mendoza}, Alexandra and {Rawls}, Meredith L. and {Plazas Malag{\'o}n}, Andr{\'e}s Alejandro},
        title = "{Identifying and Measuring Satellite Streaks in DECam Images}",
      journal = {arXiv e-prints},
         year = 2026,
        month = mar,
          eid = {arXiv:2603.10790},
        pages = {arXiv:2603.10790},
          doi = {10.48550/arXiv.2603.10790},
archivePrefix = {arXiv},
       eprint = {2603.10790},
 primaryClass = {astro-ph.IM},
       adsurl = {https://ui.adsabs.harvard.edu/abs/2026arXiv260310790S}
}

@ARTICLE{starlink_impact_halferty,
       author = {{Halferty}, Grace and {Reddy}, Vishnu and {Campbell}, Tanner and {Battle}, Adam and {Furfaro}, Roberto},
        title = "{Photometric characterization and trajectory accuracy of Starlink satellites: implications for ground-based astronomical surveys}",
      journal = {\mnras},
         year = 2022,
        month = oct,
       volume = {516},
       number = {1},
        pages = {1502-1508},
          doi = {10.1093/mnras/stac2080},
archivePrefix = {arXiv},
       eprint = {2208.03226},
 primaryClass = {astro-ph.EP},
       adsurl = {https://ui.adsabs.harvard.edu/abs/2022MNRAS.516.1502H}
}
\end{document}